\documentclass{aa}

\usepackage{graphicx}

\usepackage{caption}
\usepackage{subcaption}

\usepackage{txfonts}
\usepackage{color}

\begin{document}

\title{More insights into bar quenching. \\
 Multi-wavelength analysis of four barred galaxies}

\author{K. George\inst{1}\fnmsep\thanks{koshyastro@gmail.com},  P. Joseph\inst{2,3}, C. Mondal\inst{3}, S. Subramanian\inst{3}, A. Subramaniam\inst{3}, K. T. Paul\inst{2}}

\institute{Faculty of Physics, Ludwig-Maximilians-Universit{\"a}t, Scheinerstr. 1, Munich, 81679, Germany \and Department of Physics, Christ University, Bangalore, India \and Indian Institute of Astrophysics, Koramangala II Block, Bangalore, India}

  \abstract{The underlying nature of the process of star formation quenching in the central regions of barred disc galaxies that is due to the action of stellar bar is not fully understood. We present a multi-wavelength study of four barred galaxies using the archival data from optical, ultraviolet, infrared, CO, and HI imaging data on star formation progression and stellar and gas distribution to better understand the process of bar quenching. We found that for three galaxies,
  the region between the nuclear or central sub-kiloparsec region and the end of the bar (bar region) is devoid of neutral and molecular hydrogen. While the detected neutral hydrogen is very negligible, we note that molecular hydrogen is present abundantly in the nuclear or central sub-kiloparsec regions of all four galaxies. The bar co-rotation radius is also devoid of recent star formation for three out of four galaxies. One galaxy shows significant molecular hydrogen along the bar, which might mean that the gas is still being funnelled to the centre by the action of the stellar bar. Significant star formation is also present along the bar co-rotation radius of this galaxy. The study presented here supports a scenario in which gas redistribution as a result of the action of stellar bar clears the bar region of fuel for further star formation and eventually leads to star formation quenching in the bar region.}
  

\titlerunning{Bar-induced star formation quenching }
\authorrunning{K. George\inst{1}}

\maketitle
%

\section{Introduction}

The ongoing star formation in the central regions of disc galaxies can be affected by the action of non-axisymmetric gravitational potential exerted by stellar bars. The stellar bar can quench star formation (bar quenching) in the region between the nuclear or central sub-kiloparsec (sub-kpc) region and the end of the bar (bar region) of  spiral galaxies. A plethora of evidence supports the hypothesis that bar quenching occurs in disc galaxies in the local Universe \citep{Tubbs_1982,Reynaud_1998,Masters_2010,Masters_2012,Cheung_2013,Renaud_2013,Gavazzi_2015,Hakobyan_2016,James_2016,Cervantes_2017,Spinoso_2017,Khoperskov_2018,James_2018, George_2019a, Newnham_2020, Rosas-Guevara_2020,Krishnarao_2020,Diaz_2020}. It can be more significant at higher redshifts for which detailed nature of quenching in disc galaxies is yet to be understood from observations as well as simulations. The picture that emerged from many studies suggests that shocks or torques induced by stellar bars on the gas present in the co-rotation radius (i.e. between the central sub-kpc region and the end of the bar where stars in the disc and bar co-rotate with the same angular velocity) leads to bar quenching. However, it is also important to note that other processes such as active galactic nucleus (AGN)  feedback, morphological quenching through the action of central bulges, stellar feedback, and mergers can quench ongoing star formation in the central regions of the galaxies \citep{Martig_2009,Man_2018} and lead to global quenching of star formation.\\

The action of the stellar bar can induce torques that can accumulate gas at the periphery of the co-rotation radius. The newly accumulated gas eventually sheds the angular momentum and falls onto the central regions of the galaxy by funnelling through the stellar bar. This gas inflow leads to a range of morphologies in the distribution of gas and star formation in the central regions of barred galaxies. In the initial phase, the galaxy exhibits star formation along the bar without star formation activity in the centre, then star formation proceeds both along the bar and in the centre of the galaxy, and eventually, intense star formation occurs only in the centre of the galaxy with little star formation along the bar. The gas concentration in the central region is enhanced, leaving the bar region deficient of gas (\citealt{Combes_1985}; \citealt{Spinoso_2017}). In the absence of external gas supply, cessation of star formation occurs within the co-rotation radius or in the bar region. In this scenario of redistributed gas in the central regions of barred galaxies, an intense star formation in the central sub-kpc region is expected with a lack of neutral (HI) and  molecular hydrogen (H$_{2}$) between the nuclear or central sub-kpc and the end of the bar.\\

Another mechanism has been proposed to achieve this: the shocks induced by the stellar bar that stabilizes the gas (collected within the co-rotation radius) against collapse by increasing the turbulence. Turbulence can destroy as well as hamper the formation of giant molecular clouds. This implies that gas may be present in HI and  H$_{2}$ form within the bar region, but with very little efficiency to become converted into stars. This leads to the cessation of star formation in the bar region (\citealt{Tubbs_1982}; \citealt{Reynaud_1998}; \citealt{Verley_2007}; \citealt{Haywood_2016}; \citealt{Khoperskov_2018}).\\

The net effect of the two processes explained above is quenching of star formation within the co-rotation radius of barred galaxies. The presence or absence of gas in the bar region can provide constraints on the mechanism responsible for bar quenching in barred galaxies, however. We recently found \citep{George_2019a} based on a multi-wavelength analysis of a face-on barred galaxy, Messier 95 (also known as NGC 3351) that the bar region of Messier 95 is devoid of star formation and gas (both molecular and atomic hydrogen) and that  the central sub-kpc nuclear region shows enhanced star formation. This provided observational evidence for bar quenching through redistributed gas in the central regions of barred galaxies. The interesting question still remains is whether this phenomenon is global. Many barred galaxies are found to be devoid of recent star formation in the bar region \citep{James_2009}. Whether the process of gas redistribution by stellar bars is the dominant process for bar quenching is yet to be understood. The aim of the present study is to provide convincing evidence that gas redistribution occurs in barred disc galaxies, which can lead to bar quenching. We build on our previous work \citep{George_2019a} and extend the study with four galaxies using archival multi-wavelength data. Throughout this paper, we adopt a flat Universe cosmology with $H_{\rm{o}} = 71\,\mathrm{km\,s^{-1}\,Mpc^{-1}}$, $\Omega_{\rm{M}} = 0.27$, $\Omega_{\Lambda} = 0.73$ \citep {Komatsu_2011}.\\

\section{Data and analysis}

\begin{figure}
\centering
\begin{tabular}{cc}
    \includegraphics[width=0.42\linewidth]{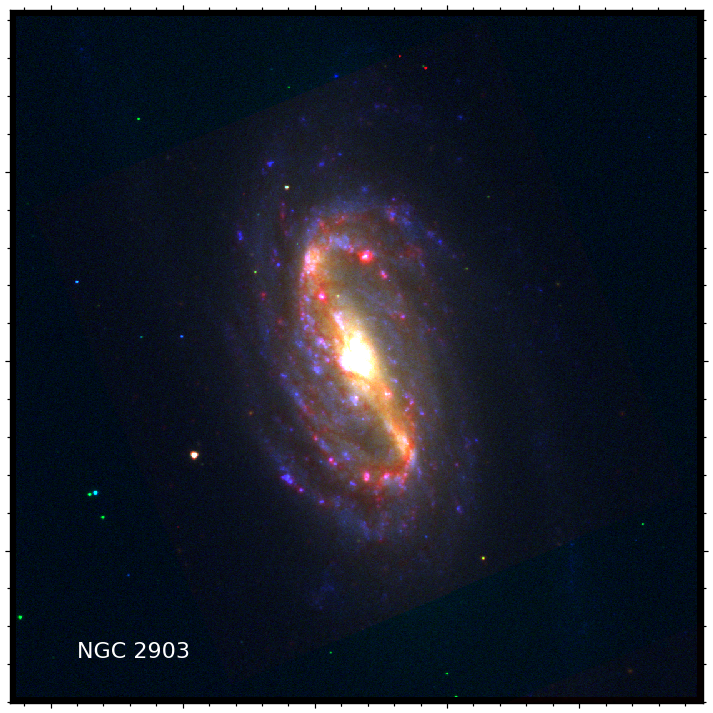}&
    \vspace{0.00mm} 
    \includegraphics[width=0.42\linewidth]{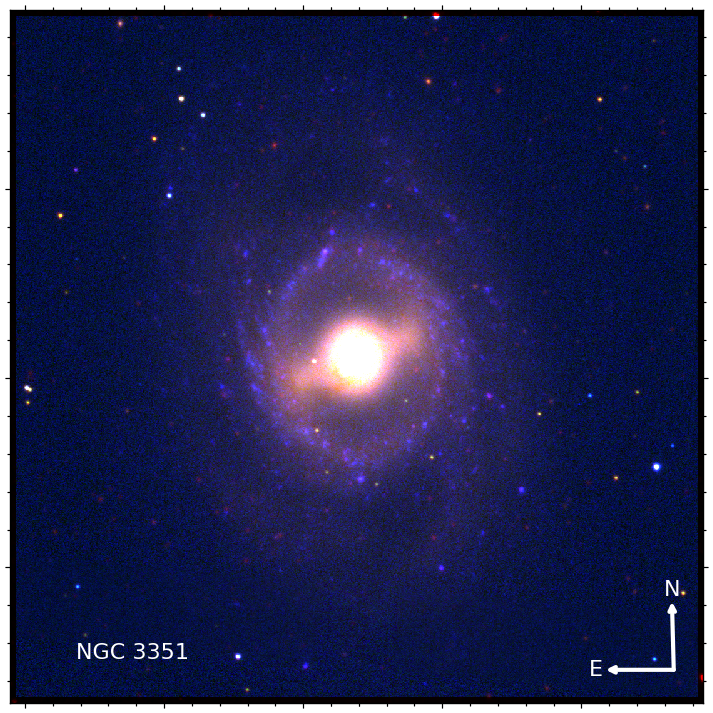}\\[2\tabcolsep]
    \vspace{0.00mm} 
    \includegraphics[width=0.42\linewidth]{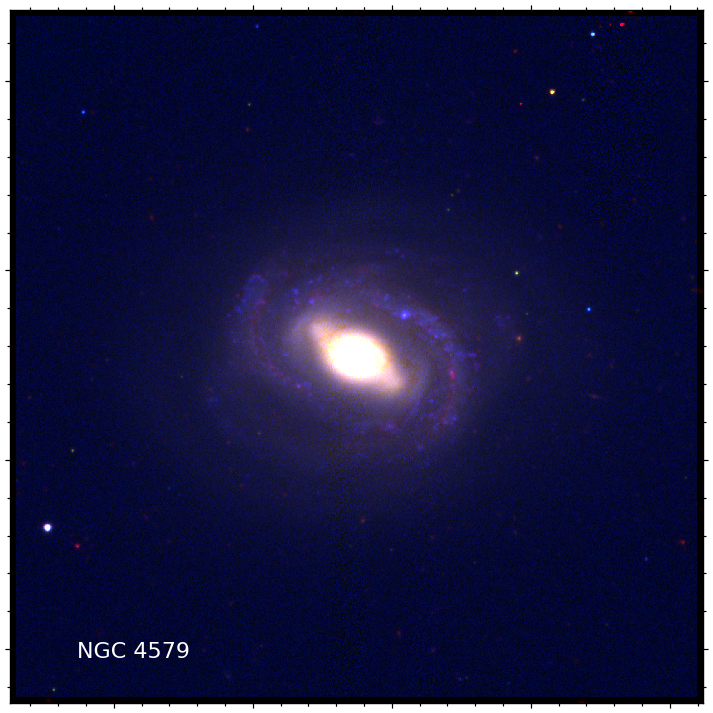}&
    \vspace{0.00mm} 
    \includegraphics[width=0.42\linewidth]{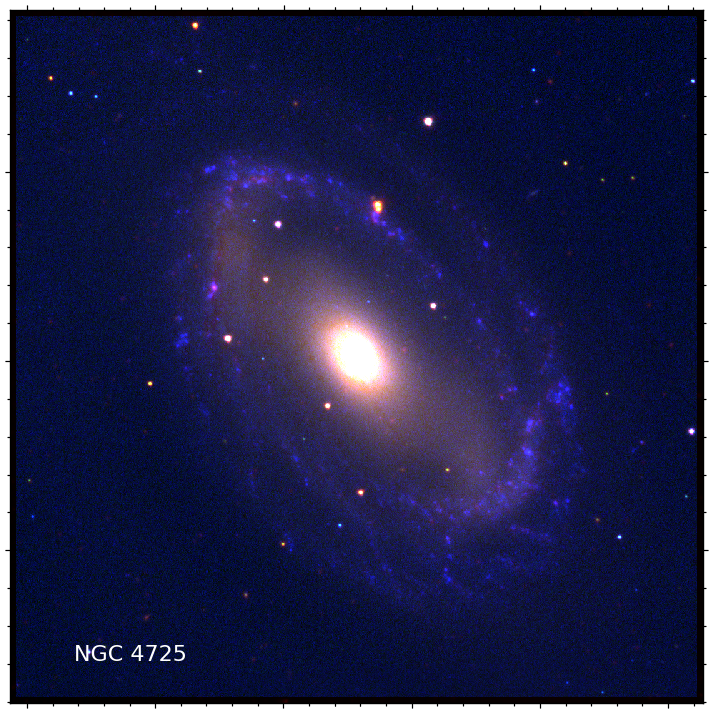}
\end{tabular}
\caption{Colour-composite images of the barred galaxies NGC 2903, NGC 3351, NGC 4579, and NGC 4725. The RGB images are created by assigning red (Spitzer 3.6 ${\mu}$ ), green (SDSS $r  \text{}$ band) and blue (SDSS $u \text{ }$band) colours to the images. All images have the same orientation; the north-east direction arrows are laid on the image of NGC 3351. 
}\label{figure:fig1}
\end{figure}

We exploited the archival multi-wavelength data of four barred galaxies (NGC 2903, NGC 3351, NGC 4579, NGC 4725) observed from ultraviolet to radio wavelengths as part of different campaigns. The samples of these four galaxies are from galaxies observed in HERA CO-Line Extragalactic Survey and Berkeley-Illinois-Maryland Association Survey of Nearby Galaxies (HERACLES; \citealt{Leroy_2009}). HERACLES used the IRAM 30 m telescope to map CO emission from 48 nearby galaxies, out of which we found that the four galaxies are strongly barred and seen nearly face-on. These four galaxies were selected purely based on the availability of multi-wavelength data that  span from radio to ultraviolet wavelengths and trace different components of the interstellar medium and the underlying stellar population.
The CO line emission traces  H$_{2}$ in galaxies, and CO maps can be used to determine the H$_{2}$ distribution in galaxies (see \citet{Young_1991}).
In the next section we describe  the basic properties of the four galaxies. The Sloan Digital Sky Survey (SDSS) $uz$ DR12 (\citealt{Alam_2015}) optical imaging data along with the Spitzer IRAC 3.6 ${\mu}$ image of galaxies was used to construct the colour-composite image shown in Figure~\ref{figure:fig1}. The optical and infrared images show a prominent stellar bar, and blue ($u$) and red (3.6 ${\mu}$) images help visualise the spatial variation in the relative contribution of the young and evolved stellar populations in the galaxy.\\

All the four galaxies are observed in the far-ultraviolet (FUV) $\lambda_{eff}$=1538.6 {\AA} and near-ultraviolet (NUV) $\lambda_{eff}$=2315.7 {\AA} wavelengths using the NASA GALEX mission \citep{Martin_2005}. The GALEX FUV channel imaging is at $\sim$ 4.2" and the NUV channel imaging is at 5.3" resolution \citep{Morrissey_2007}. The FUV image was degraded to NUV resolution by running a Gaussian 2D kernel of width 0.57". The GALEX GR6/GR7 data were pipeline reduced (with good photometric quality) and astrometry calibrated.  The HI maps for two galaxies (NGC 2903 and NGC 3351) are taken from The HI Nearby Galaxy Survey (THINGS; \citealt{Walter_2008}). The HI maps for NGC 4579 and  NGC 4725 are taken from archival Very Large Array (VLA) data (provided by Adam Leroy, private communication). The CO map (J$_{2-1}$ transition) from CO was measured as part of HERACLES \citep{Leroy_2009}. The infrared image from the Spitzer IRAC 3.6 $\mu$ channel was observed as part of Spitzer Survey of Stellar Structure in Galaxies (S$^{4}$G) \citep{Sheth_2010}. \\

The foreground extinctions from the Milky Way galaxy in the direction of the four galaxies are  $A_{V}$ = 0.085 for NGC 2903, $A_{V}$ = 0.076 for NGC 3351, $A_{V}$ = 0.112 for NGC 4579, and $A_{V}$ = 0.032 for NGC 4725 (taken from the NED; \citet{Schlegel_1998}), which we scaled to the FUV and NUV $\lambda_{mean}$ values using the \citet{cardelli_1989} extinction law. We also corrected the magnitudes.  The regions of the FUV and NUV images that correspond to the galaxy were then isolated using the background counts from the whole image set as a threshold. Pixels with values above the 3$\sigma$ of the threshold were selected to isolate the galaxy. The counts in the selected pixels were background subtracted, integration time weighted, and converted into magnitude units using the zero-points of \citet{Morrissey_2007}. Magnitudes for each pixel were used to compute the FUV$-$NUV colour map of the galaxy. The pixels were colour-coded in units of FUV$-$NUV colour. \\

The FUV$-$NUV colour map can be used to understand the star formation history of galaxies, and in particular, it can offer insights into the last burst of star formation. We used the {\tt Starburst99} stellar synthesis code to characterise the age of the underlying stellar population in the four galaxies \citep{Leitherer_1999}. We selected 19 spectra of different ages (in an interval of 10 Myr between 1 - 100 Myr and an interval of 100 Myr between 100 - 900 Myr) from the Starburst99 single stellar population (SSP) models assuming a Kroupa initial mass function (IMF)  \citep{Kroupa_2001} and solar metallicity (Z=0.02). The synthetic spectral energy distribution (SED) for a given age was then convolved with the effective area of the FUV and NUV pass-bands to compute the expected fluxes. The estimated values were then used to calculate the SSP ages corresponding to the observed FUV$-$NUV colours.  We performed a linear interpolation for the observed colour value and estimated the corresponding ages in all pixels in the FUV$-$NUV colour map. The error in the estimated ages mainly arises from the photometric error of the (FUV-NUV) colour in each pixel. We considered the photometric error and estimated the corresponding error in age with the same technique. The FUV and NUV flux is subjected to extinction at the rest-frame of the galaxy. We do not have a proper extinction map of the galaxies studied here. We also note that the ages are estimated for solar metallicity. The FUV$-$NUV pixel colour maps and the derived ages can therefore be considered as the upper limits of the actual values.\\

The Spitzer IRAC 3.6 ${\mu}$ image of a galaxy can be used as an extinction-free tracer for the evolved stellar population that dominates the underlying stellar mass \citep{Meidt_2014}. We extracted the contour over the stellar bar from the Spitzer IRAC 3.6 $\mu$ image of the galaxies to identify the extent of the bar that is dominated by evolved  stellar population and to overlay this information on other wavelength images.

\begin{table*}
\centering
\label{galaxy details}
\tabcolsep=0.1cm
\begin{tabular}{clccccc} 
\hline
Name & Morphology & Redshift & log axis ratio & Major-axis position angle & log Stellar mass & Metallicity \\
 & & $z$ &  & $\deg$ & M$_{\odot}$  & 12 $+$ logO/H \\
\hline
NGC 2903 & SBd  & 0.0018 & 0.35  & 22.0  &10.665 & 8.90\\
NGC 3351 & SBb  & 0.0026 & 0.21  & 10.7  &10.486 &  8.90\\
NGC 4579 & SBb  & 0.0051 & 0.12  & 90.2  &11.098 &  8.88\\
NGC 4725 & SBab & 0.0040 & 0.14  & 35.7  &10.881  & 8.73\\
\hline
\end{tabular}
\caption{\label{t7} Details of the four barred galaxies. Morphology and gas-phase metallicity are taken from \citet{Schruba_2011}. The stellar mass is taken from \citet{Sheth_2010}. Redshift information is taken from the Nasa/IPAC Extragalactic Database. The log of the major axis /minor axis and the major-axis position angle is measured north-eastwards based on information from HyperLeda.}
\end{table*}

The basic details of the four galaxies are given in Table 1. The stellar masses of all four galaxies are very similar, as measured from the 3.6 ${\mu}$ Spitzer imaging data. The gas-phase metallicities of the galaxies are slightly higher than solar (12$+$log O/H=8.69) \citep{Asplund_2009}.  We summarise the properties of each galaxy below, along with details on the observations and analysis.

\subsection{NGC 2903}

NGC 2903 ($\alpha$ (J2000) = 09:32:10.1 and $\delta$(J2000) = $+$21:30:03) is a barred galaxy that could be hosting a young bar with an age between 200-600 Myr with a strong inflow of gas towards the centre along the bar \citep{Leon_2008}. The galaxy is seen nearly face-on, as shown in Figure~\ref{figure:fig1}. The galaxy is non-interacting and isolated \citep{Irwin_2009}, with many intense emission regions (“hot spots”) in the nuclear region along with a ring of star formation \citep{Perez_2000}. This galaxy belongs to the field, according to the identification of barred galaxies and environment by \citet{vandenberg_2002}.
High-resolution imaging observation with HST/NICMOS revealed that the hot spots are groups/individual young stellar clusters \citep{Alonso_2001}. The integrated star formation rate of the galaxy is 2.2 M$_{\odot}$/yr \citep{Irwin_2009}. Significant H$\alpha$ and CO flux is detected along the bar of this galaxy \citep{Popping_2010}; the CO emission is more concentrated along the bar \citep{Regan_1999}.\\

\begin{figure*}[h]
\centering
\begin{tabular}{ccc}
    \includegraphics[width=5.7cm,height=5.7cm,keepaspectratio]{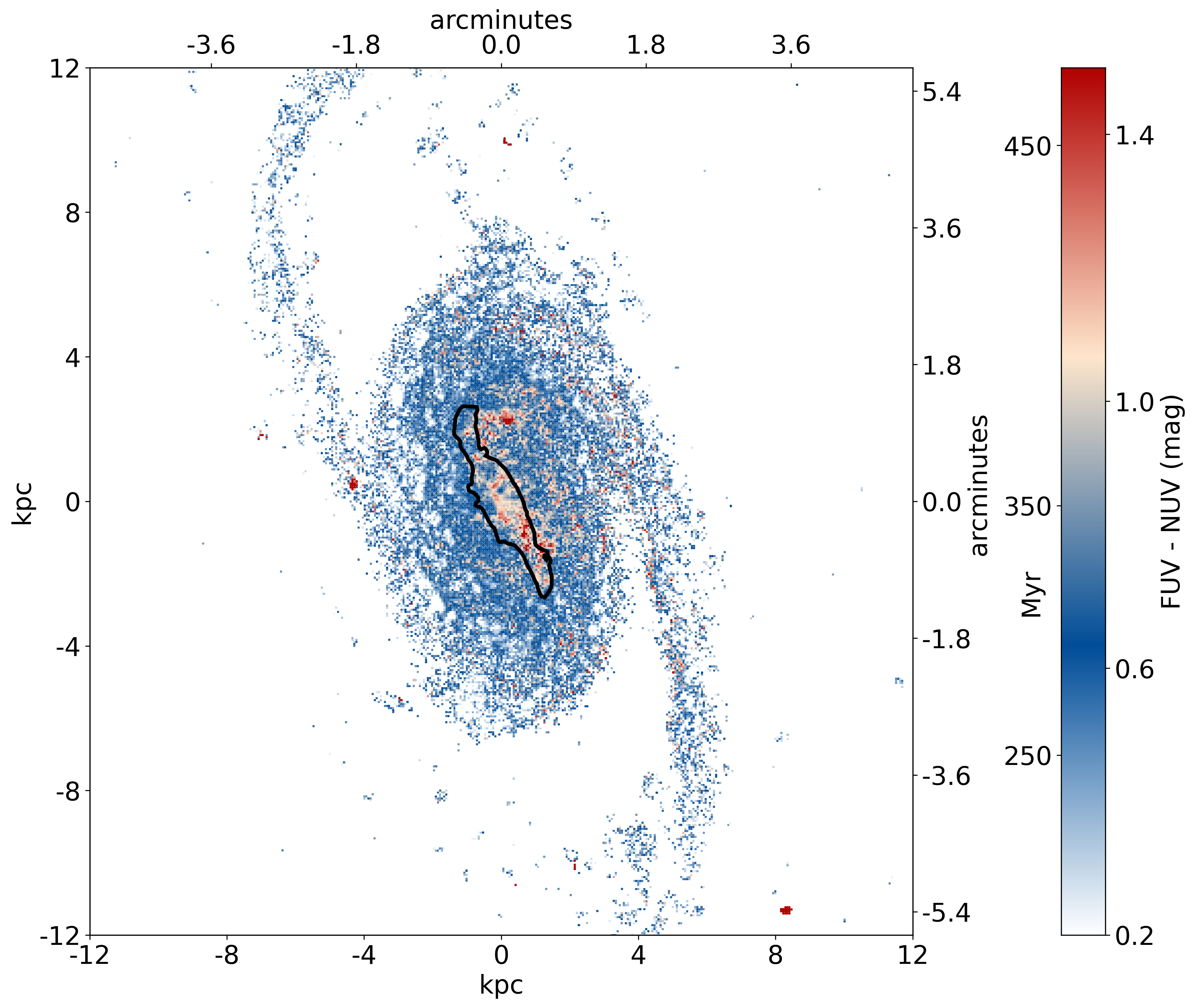}&
    \includegraphics[width=5.7cm,height=5.7cm,keepaspectratio]{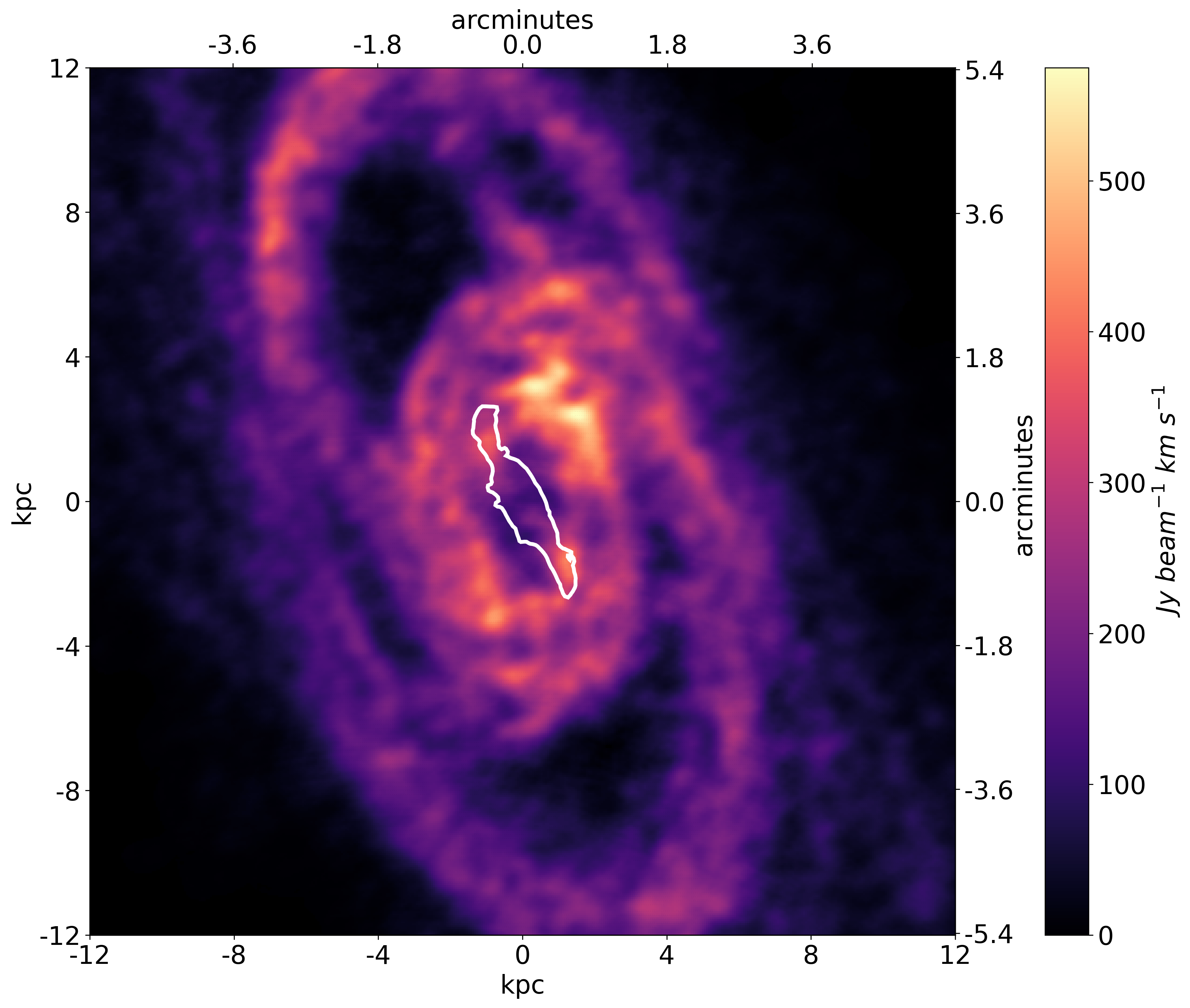}&
    \includegraphics[width=5.7cm,height=5.7cm,keepaspectratio]{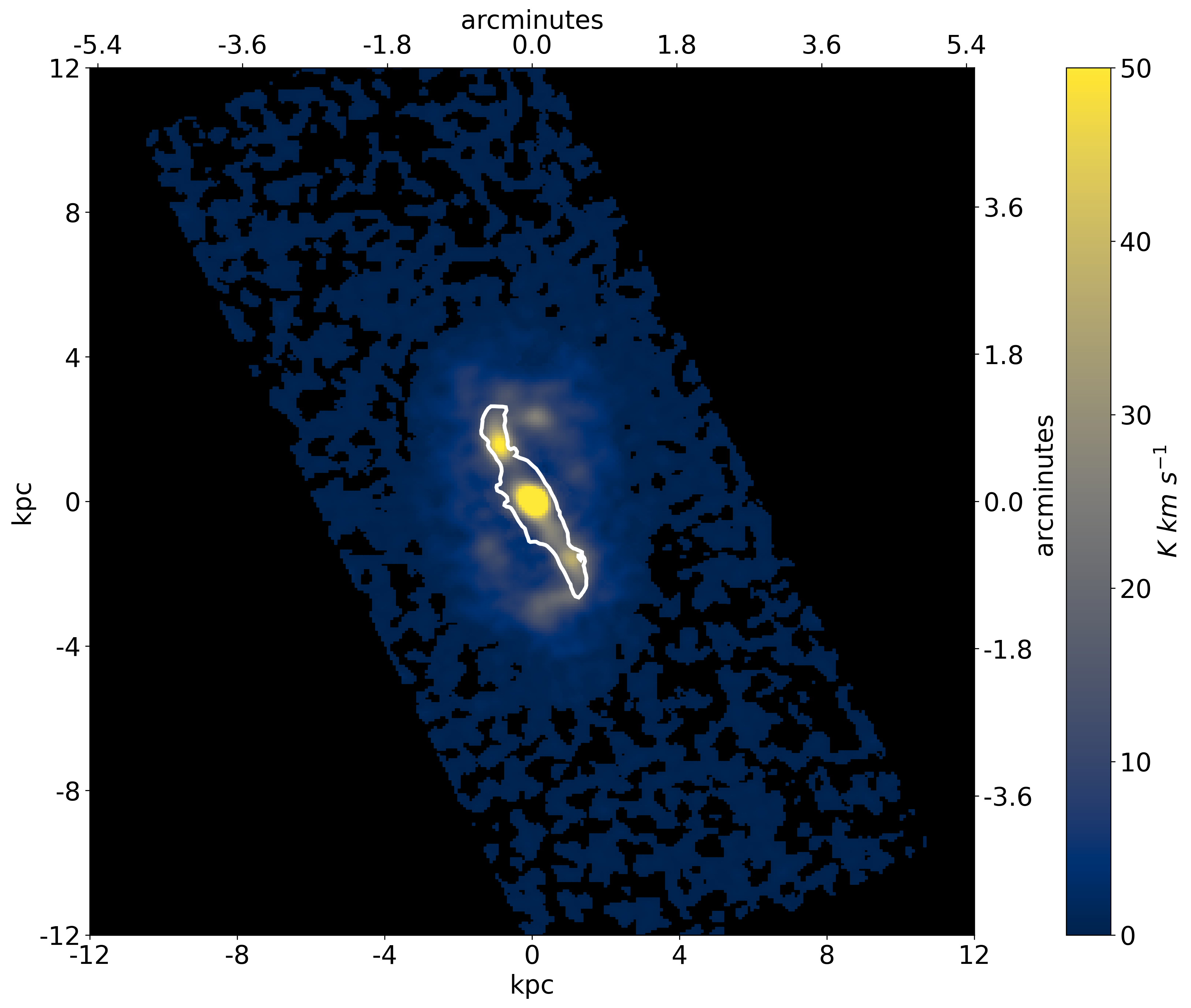}\\[2\tabcolsep]
    \includegraphics[width=5.7cm,height=5.7cm,keepaspectratio]{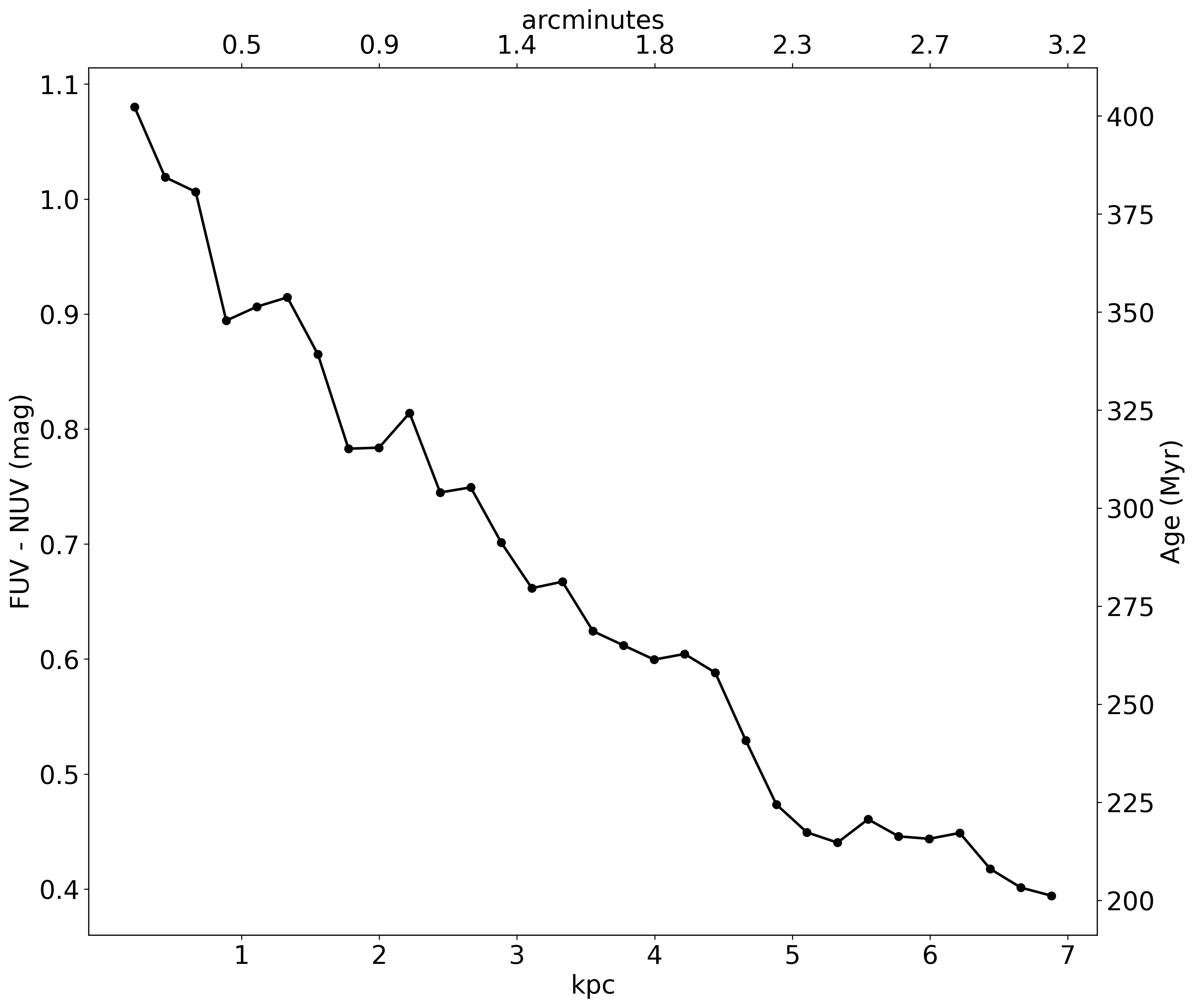}&
     \includegraphics[width=5.7cm,height=5.7cm,keepaspectratio]{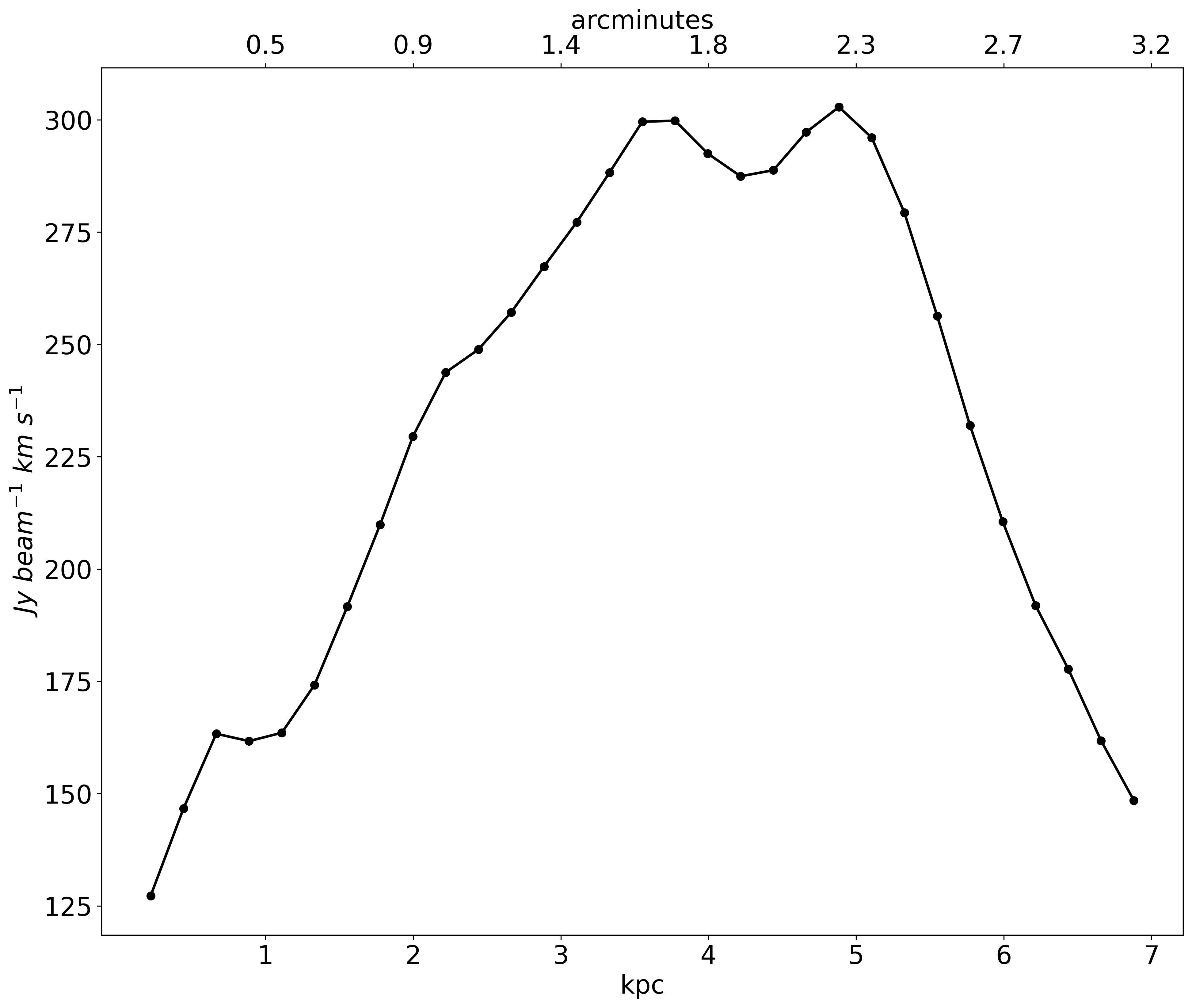}&
      \includegraphics[width=5.7cm,height=5.7cm,keepaspectratio]{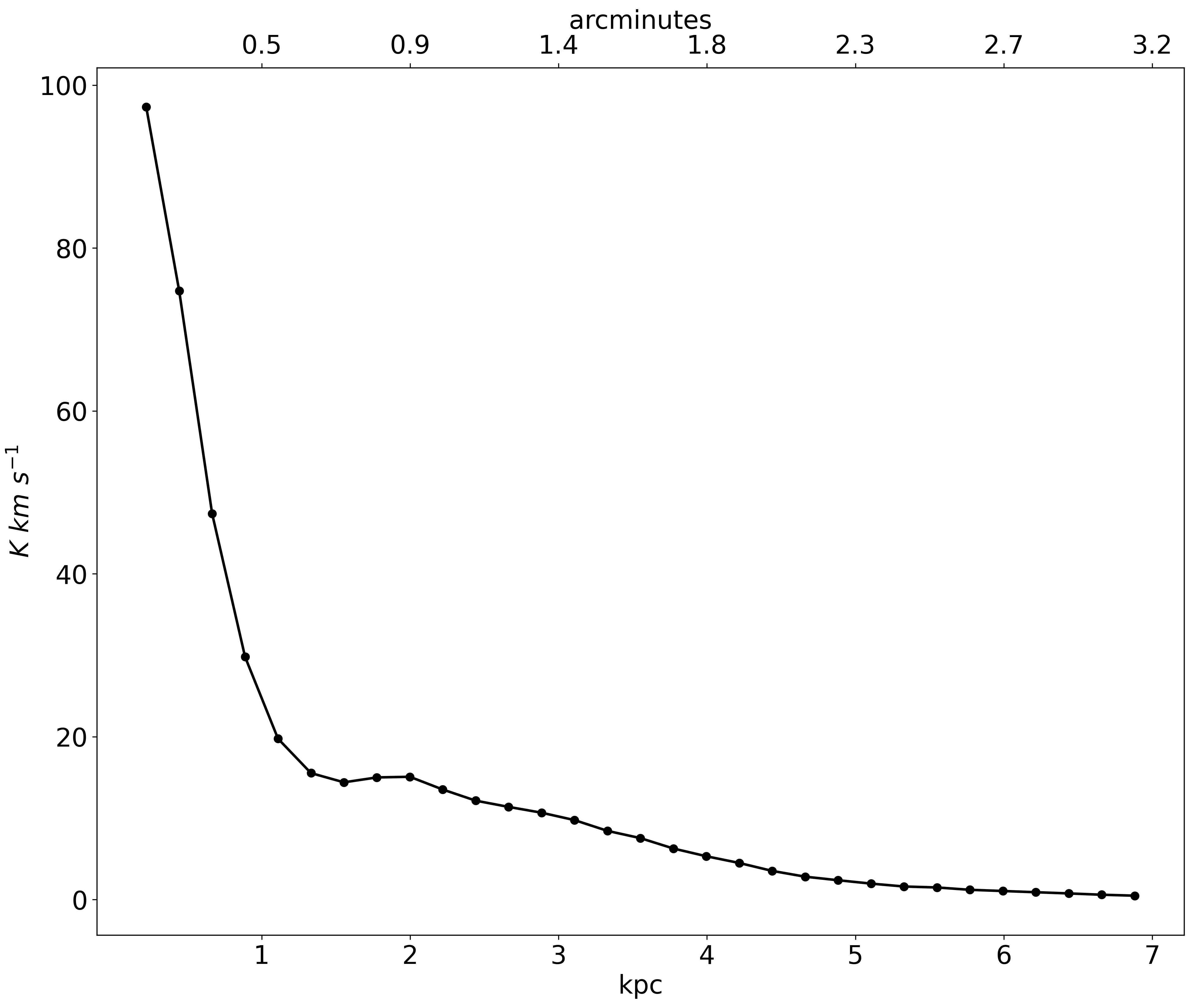}
\end{tabular}
\caption{NGC 2903. First row: 1) $FUV-NUV$ colour map. The black contour shows the stellar bar detected in the Spitzer IRAC 3.6 $\mu$ galaxy image. The pixels are colour-coded in units of $FUV-NUV$ colour. The corresponding SSP equivalent ages are also noted in the colour bar. 2) The HI and H$_2$ galaxy maps are shown in the next two columns; the stellar bar is plotted in white. The corresponding azimuthally averaged profile of the galaxy is shown in the bottom row. The main bodies of the galaxies were averaged in elliptical annuli of width 0.1 arcmin (0.22 kpc) taking the axis ratio and position angle of the galaxy into account.}\label{figure:fig2}
\end{figure*}

Figure~\ref{figure:fig2} shows the maps and profiles we created to understand the FUV$-$NUV colour and the HI and  H$_{2}$ (as traced by CO) distribution of NGC 2903. The FUV$-$NUV colour map of the galaxy displays a redder region along the bar and a blue region all around the bar. The FUV$-$NUV colour profile shows that the region near the bar is bluer than the central regions. This exercise shows that the region along the major axis of the bar hosts stellar populations of age $\geq$ 350 Myr and the region around the bar hosts very recent star formation ($\sim$ 150-250 Myr). We also note that the optical colour-composite image of this galaxy in Figure~\ref{figure:fig1} shows features along the bar that are indicative of dust lanes. This might be the reason for the redder regions along the bar and slightly beyond. The HI map and profile show a reduced HI content at the centre, along the bar, and in the region close to it compared to the region outside the bar. The molecular hydrogen map and profile clearly show significant CO emission along the bar and at the centre. The combined FUV$-$NUV and HI and H$_2$ map analysis reveals a reduction in HI distribution in the central regions, but significant star formation and  H$_{2}$ (along the bar and at the centre) within the co-rotation radius of the bar.

\subsection{NGC 3351}

NGC 3351 ($\alpha$(J2000) = 10:43:57.7 and $\delta$(J2000) = $+$11:42:14) (also known as Messier 95) is an early-type barred spiral galaxy in the Leo I group. The galaxy has an HI mass, H$_2$ mass, and integrated star formation rate of $\sim$ 10$^{9.2}$ M$_{\odot}$, $\sim$ 10$^{9}$ M$_{\odot}$ , and $\sim$ 0.940 M$_{\odot}$/yr, respectively \citep{Leroy_2008}. It is a nearly  face-on galaxy  with a prominent bar (see Figure~\ref{figure:fig1}). This galaxy belongs to the cluster in the identification of barred galaxies and environment  by \citet{vandenberg_2002}. It shows nuclear star formation and hosts a star-forming circumnuclear ring with a diameter of $\sim$ 0.7 kpc. This sub-kpc scale star formation is well studied in X-rays (\citealt{Swartz_2006}), UV (\citealt{Ma_2018}; \citealt{Colina_1997}), H$\alpha$ (\citealt{Planesas_1997,Bresolin_2002}), and in the near-infrared (\citealt{Elmegreen_1997}). In a  multi-wavelength study from the UV to the mid-infrared of the nuclear ring of NGC 3351, \citet{Ma_2018} presented the integrated properties of the ring and their correlation with bar strength. \cite{Mazzalay_2013} presented the properties of molecular gas within $\sim$ 300 pc of this galaxy using near-infrared integral field spectrograph SINFONI  at the Very Large Telescope and suggested that the nuclear region might host a stellar population of a few million years. H$\alpha$ imaging of the larger area of NGC 3351 shows that the bar region is devoid of emission (\citealt{James_2009}).\\

\begin{figure*}[h]
\centering
\begin{tabular}{ccc}
    \includegraphics[width=5.7cm,height=5.7cm,keepaspectratio]{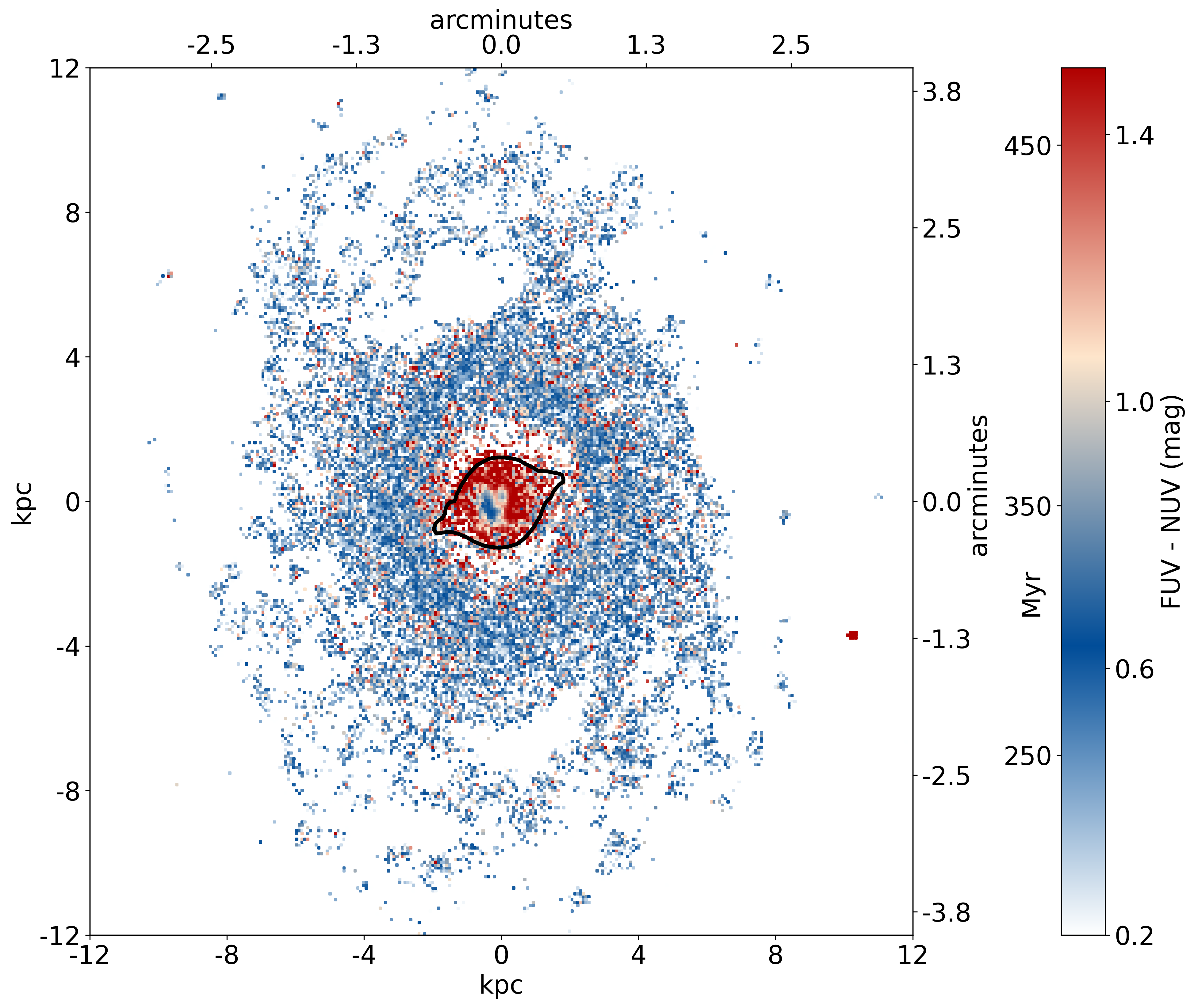}&
    \includegraphics[width=5.7cm,height=5.7cm,keepaspectratio]{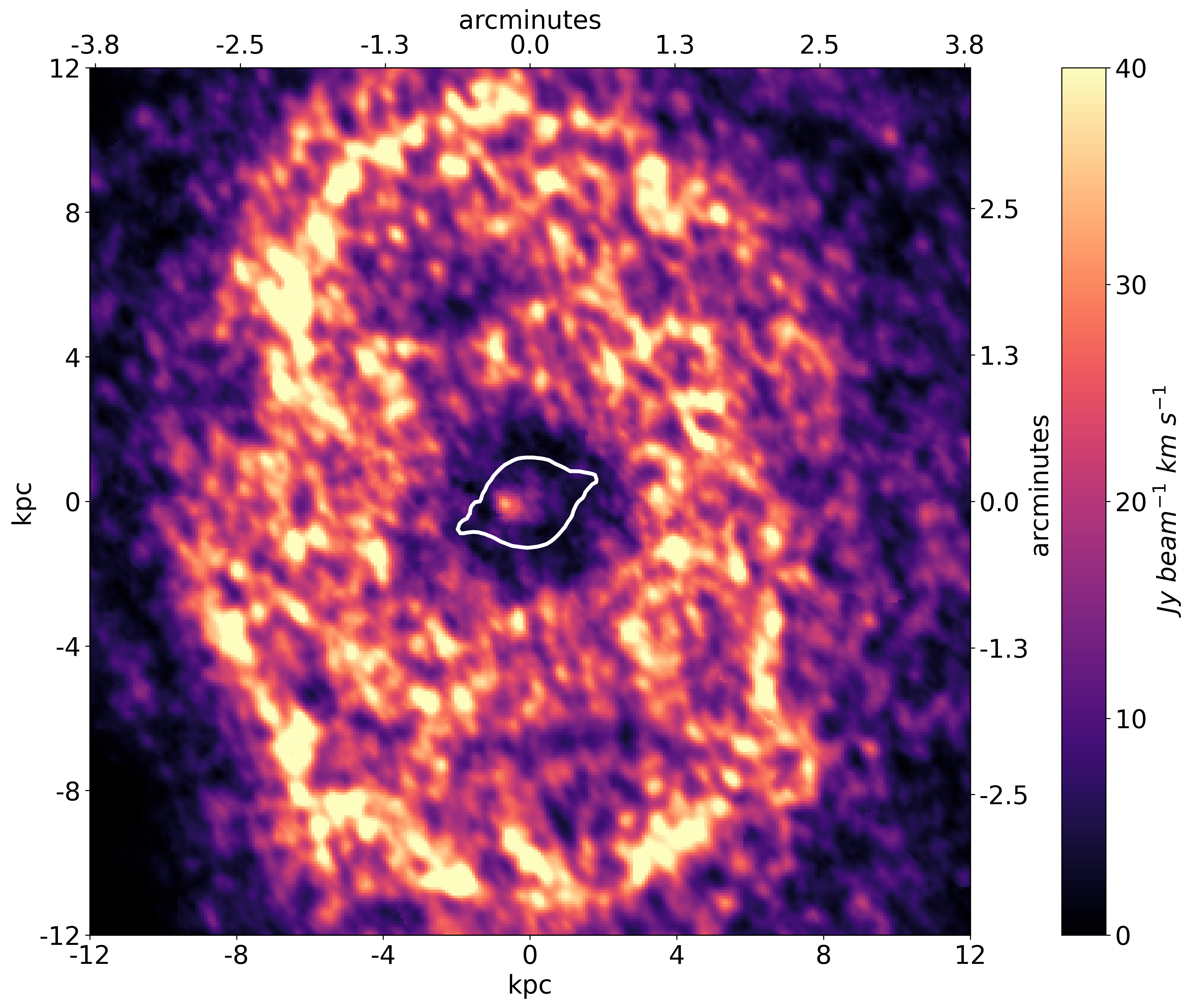}&
    \includegraphics[width=5.7cm,height=5.7cm,keepaspectratio]{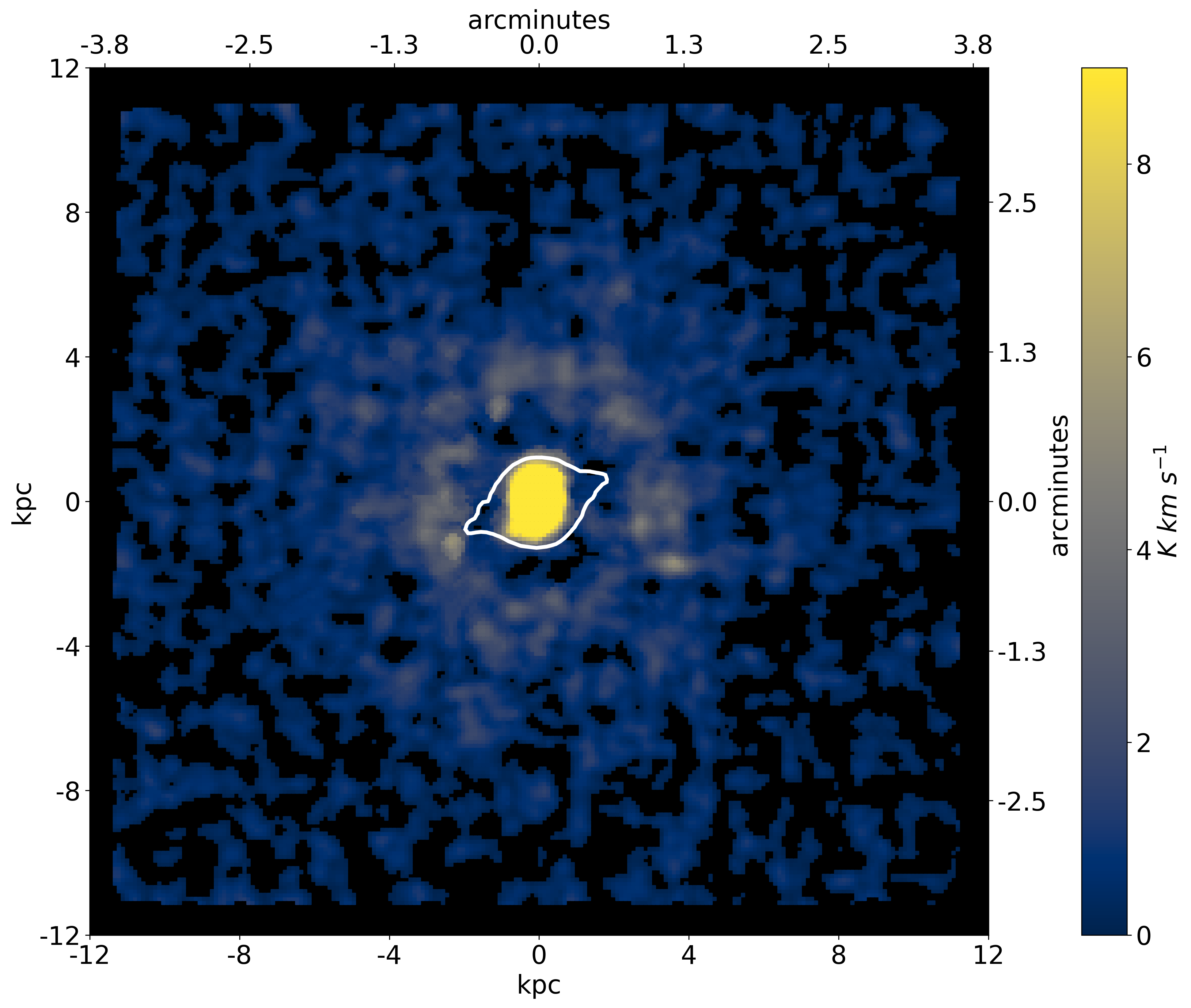}\\[2\tabcolsep]
    \includegraphics[width=5.7cm,height=5.7cm,keepaspectratio]{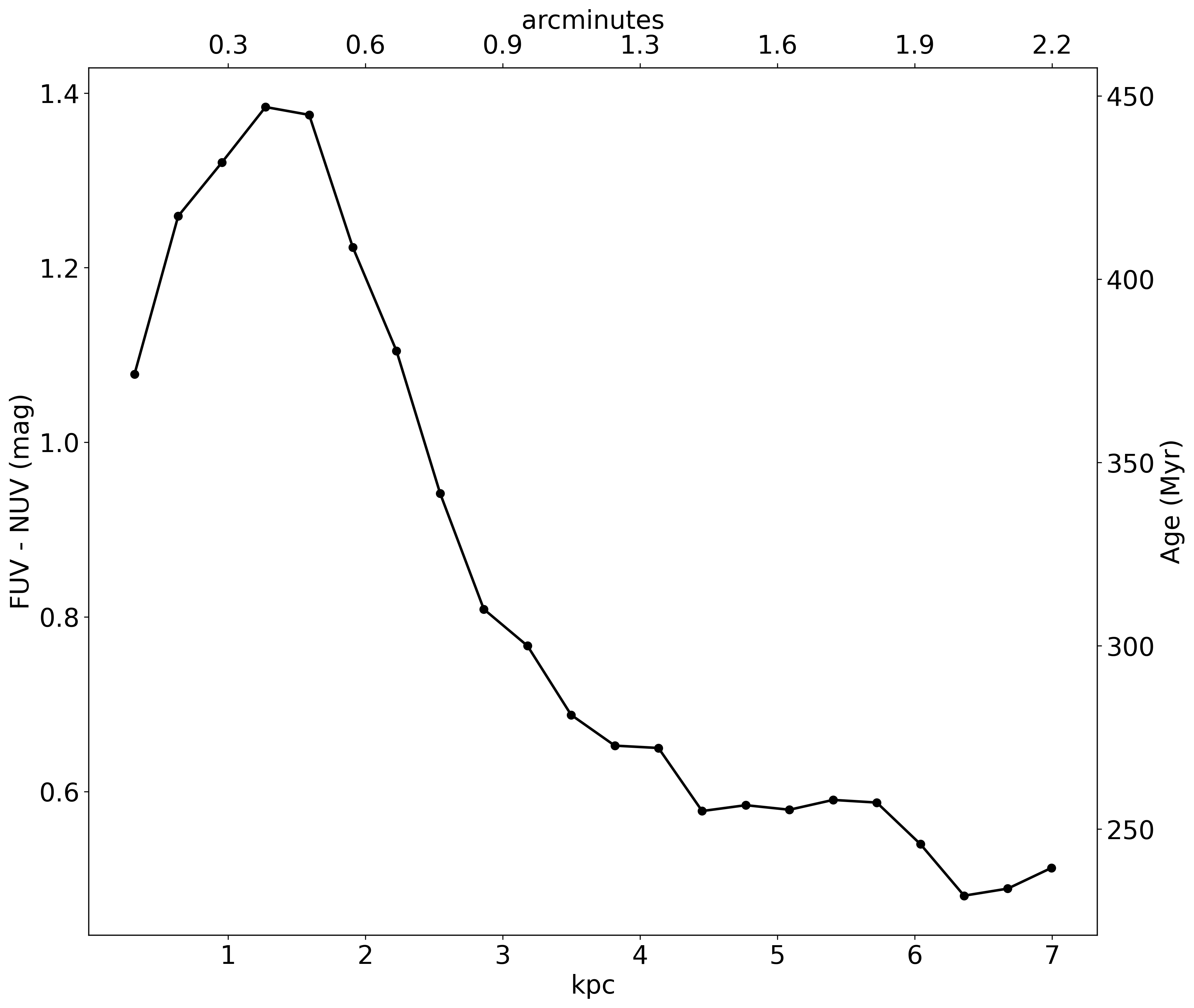}&
     \includegraphics[width=5.7cm,height=5.7cm,keepaspectratio]{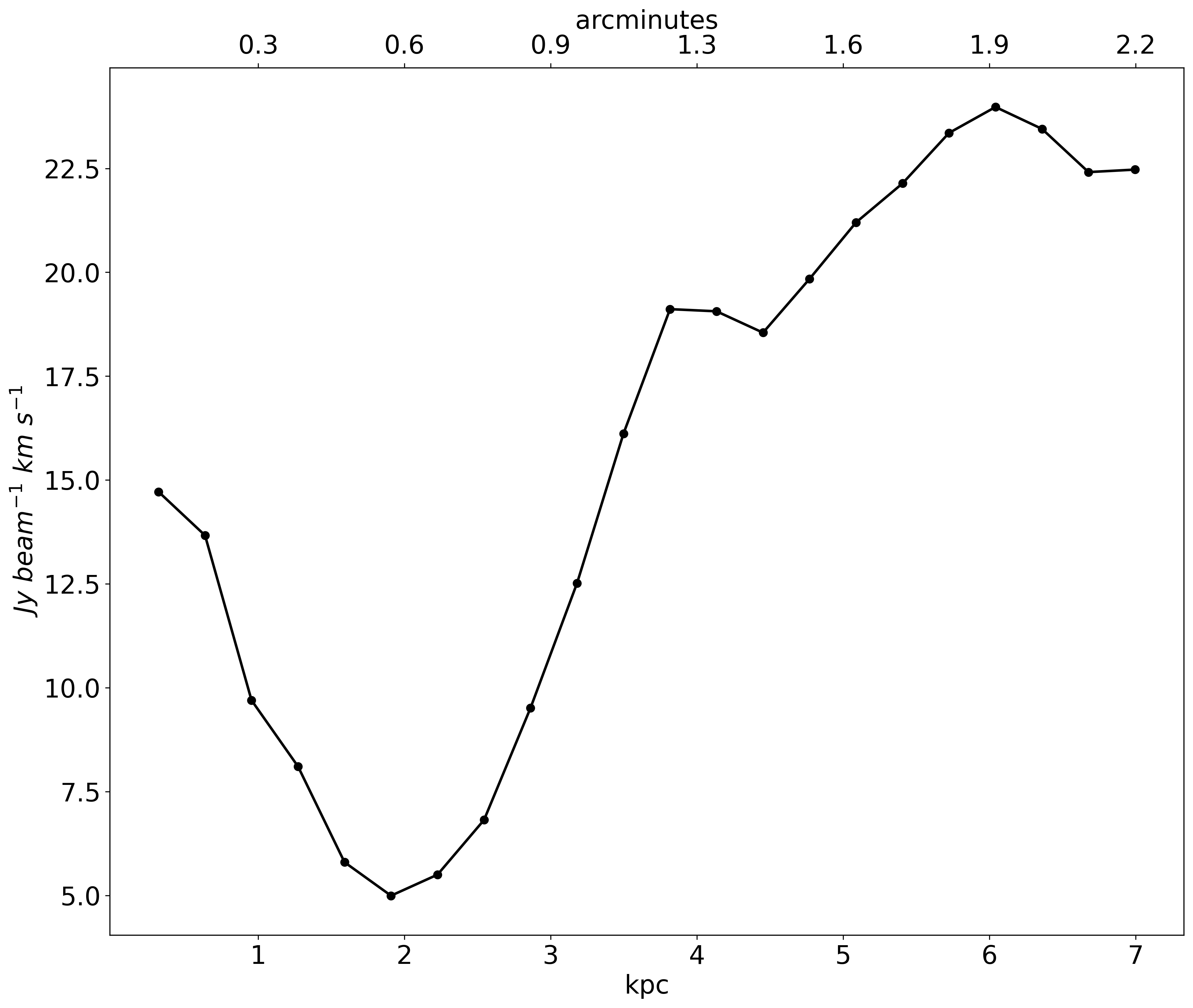}&
      \includegraphics[width=5.7cm,height=5.7cm,keepaspectratio]{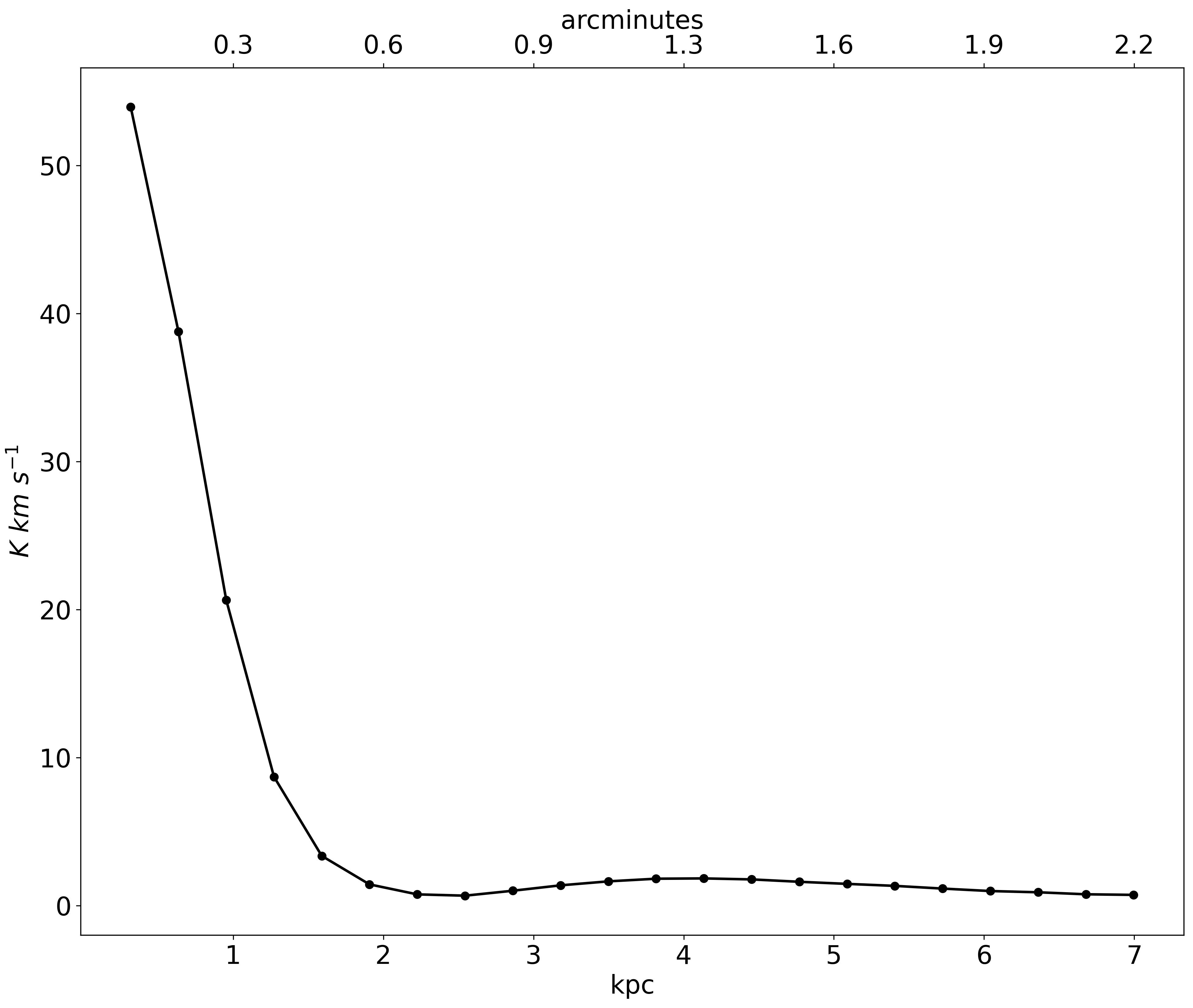}
\end{tabular}
\caption{NGC 3351. First row: 1) $FUV-NUV$ colour map. The black contour shows the stellar bar detected in the Spitzer IRAC 3.6 $\mu$ galaxy image. The pixels are colour-coded in units of $FUV-NUV$ colour. The corresponding SSP equivalent ages are also noted in the colour bar. 2) The HI and CO galaxy maps are shown in the next two columns; the stellar bar is plotted in white. The corresponding azimuthally averaged profile of the galaxy is shown in the bottom row. The main bodies of the galaxies were averaged in elliptical annuli of width 0.1 arcmin (0.32 kpc) taking the ellipticity and position angle of the galaxy into account.}\label{figure:fig3}
\end{figure*}

Figure~\ref{figure:fig3} shows the maps and profiles we created to understand the FUV-NUV colour and HI and  H$_{2}$ distribution of NGC 3351. The FUV--NUV colour map and profile of the galaxy display a redder region at the centre (with an embedded small blue clump) that is separated from the rest of the galaxy by a region with negligible UV flux. The redder region coincides with the major axis of the bar. The region around the bar has negligible UV flux. This region also coincides with the region that was identified to be devoid of emission in $H\alpha$ \citep{James_2009}. This exercise shows that the region along the major axis of the bar hosts stellar populations of age $\geq$ 350 Myr and the nuclear or central sub-kpc region shows an embedded bluer and younger clump of star formation ($\sim$ 150-250 Myr). When we compare the FUV--NUV colour map with the HI and H$_2$ maps, the circular region (the length covered by the stellar bar), without the central nuclear region, lacks molecular/neutral hydrogen and star formation. The region outside the bar hosts significant HI, and  H$_{2}$ is concentrated in a ring, as the profiles show. The central sub-kpc nuclear region of the galaxy hosts a significant molecular gas content, star formation, and some amount of HI.\\

\subsection{NGC 4579}

NGC 4579 ($\alpha$(J2000) = 12:37:43.5 and $\delta$(J2000) = $+$11:49:05)  (also known as Messier 58) is a barred spiral galaxy belonging to the Virgo cluster. The galaxy is seen nearly face-on,  as shown in Figure~\ref{figure:fig1}. This galaxy belongs to the cluster in the identification of barred galaxies and environment by \citet{vandenberg_2002}.
The galaxy is known to have a low-ionization nuclear emission-line region (LINER) nucleus with broad wings in the H$\alpha$ emission-line profile \citep{Stauffer_1982, Keel_1983, Filippenko_1985}, indicating the presence of an AGN (Seyfert 1.9) at the centre \citep{Ho_1997a,Panessa_2002}. The flux from the nuclear region can also be composed of the contribution from a starburst along with the presence of AGN \citep{Contini_2004}. Based on a kinematic analysis of face-on barred galaxies in the local Universe, \citet{Gadotti_2005} argued that the stellar bar of NGC 4579 has formed relatively recently.

\begin{figure*}[h]
\centering
\begin{tabular}{ccc}
    \includegraphics[width=5.7cm,height=5.7cm,keepaspectratio]{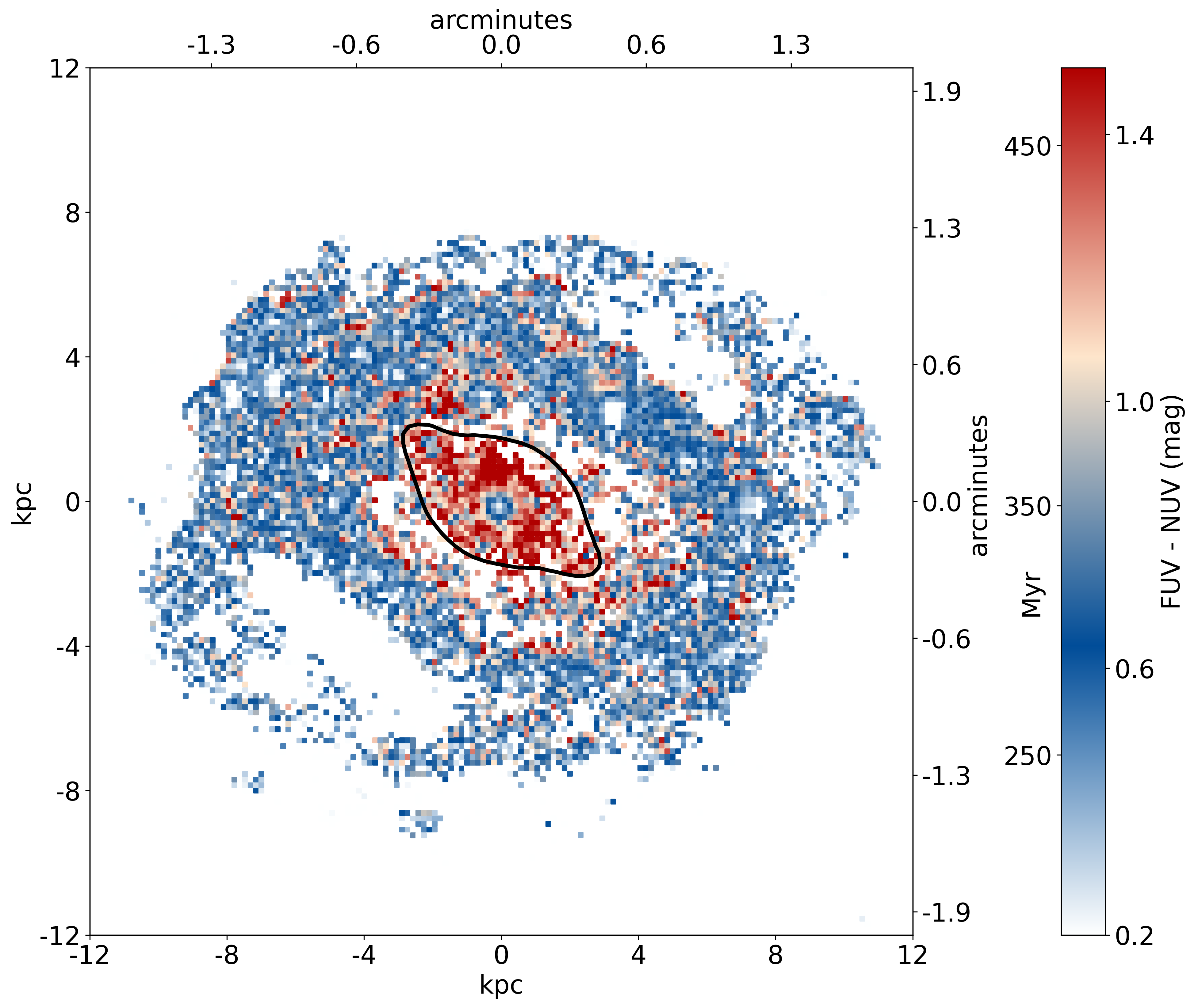}&
    \includegraphics[width=5.7cm,height=5.7cm,keepaspectratio]{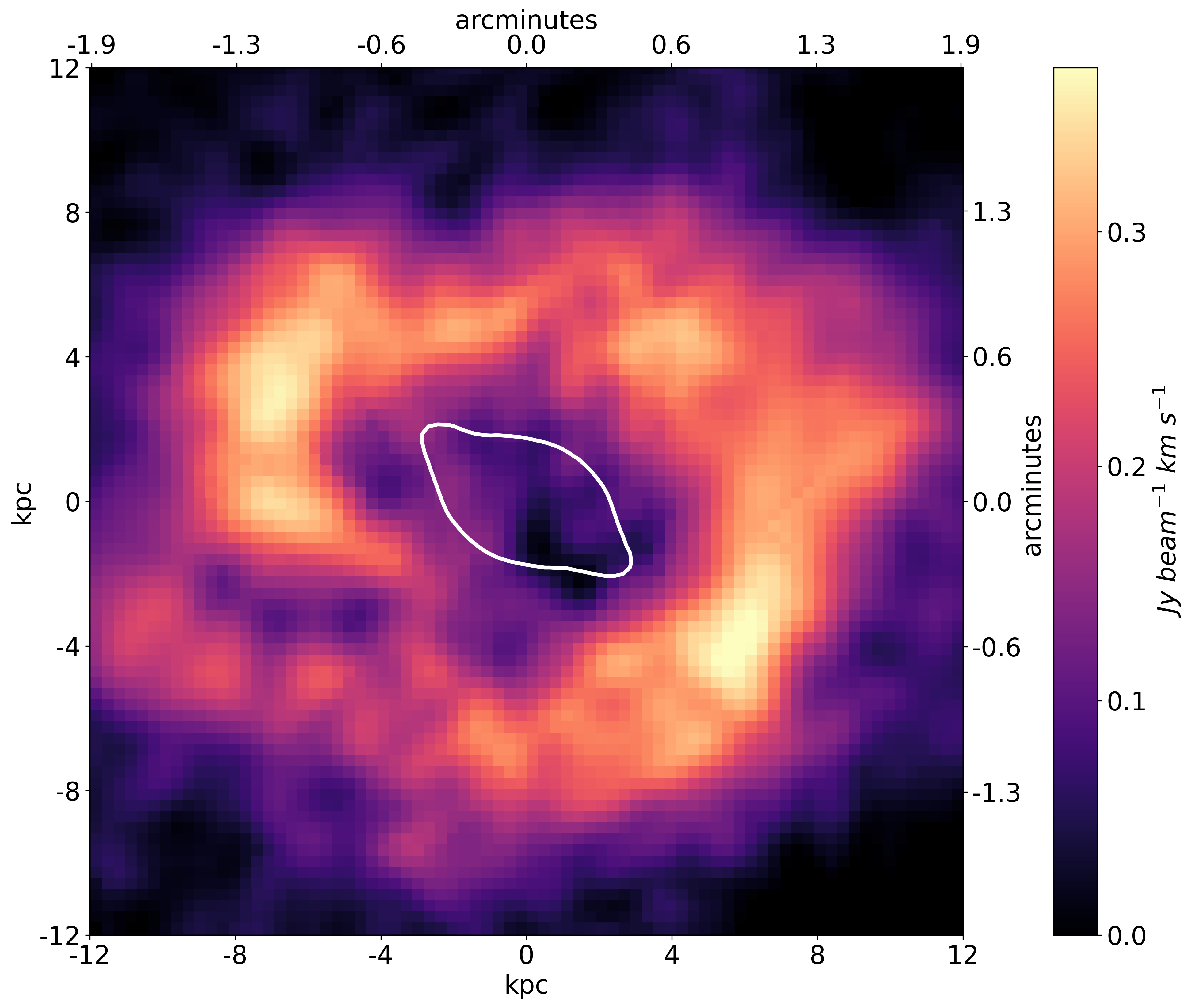}&
    \includegraphics[width=5.7cm,height=5.7cm,keepaspectratio]{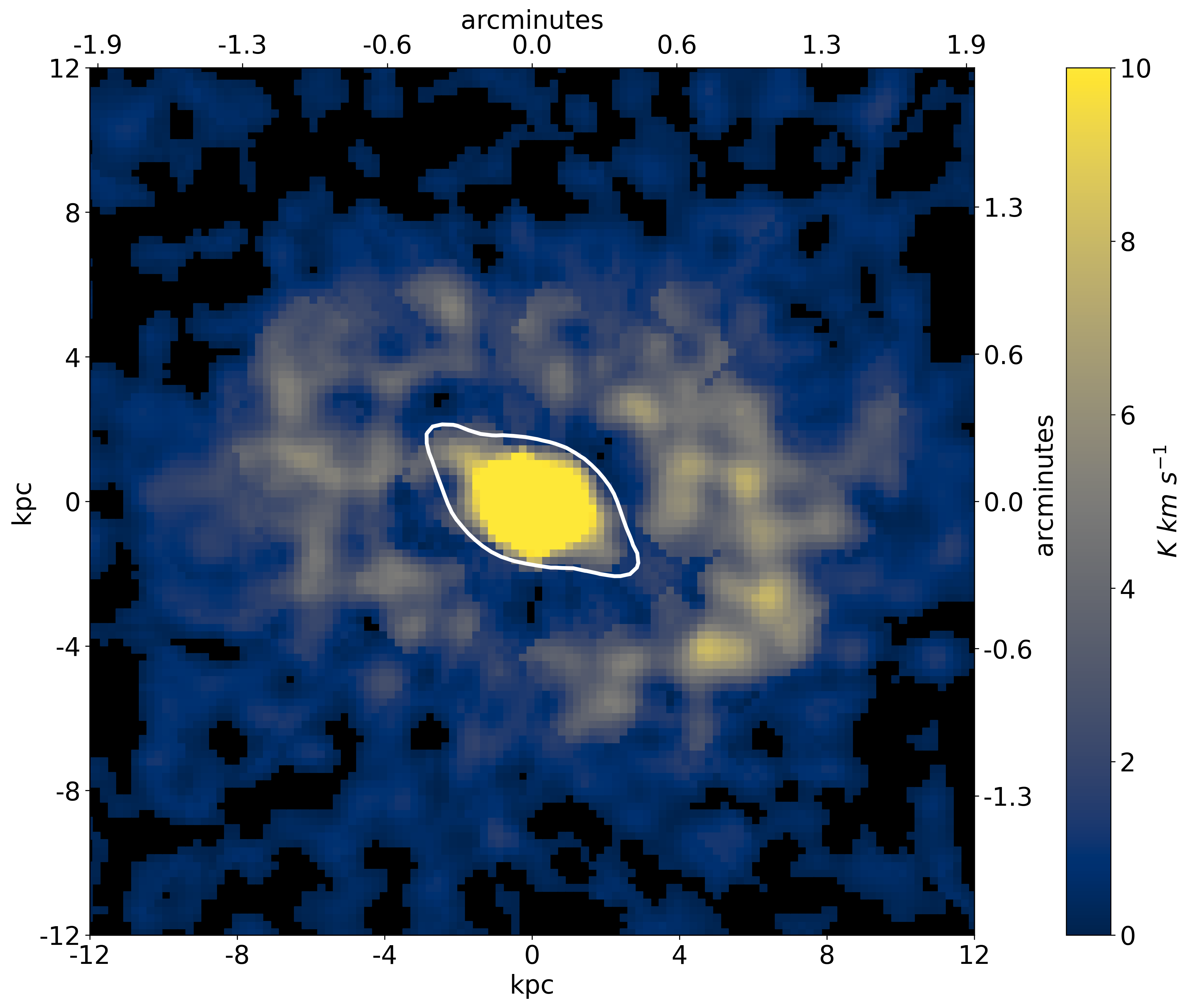}\\[2\tabcolsep]
    \includegraphics[width=5.7cm,height=5.7cm,keepaspectratio]{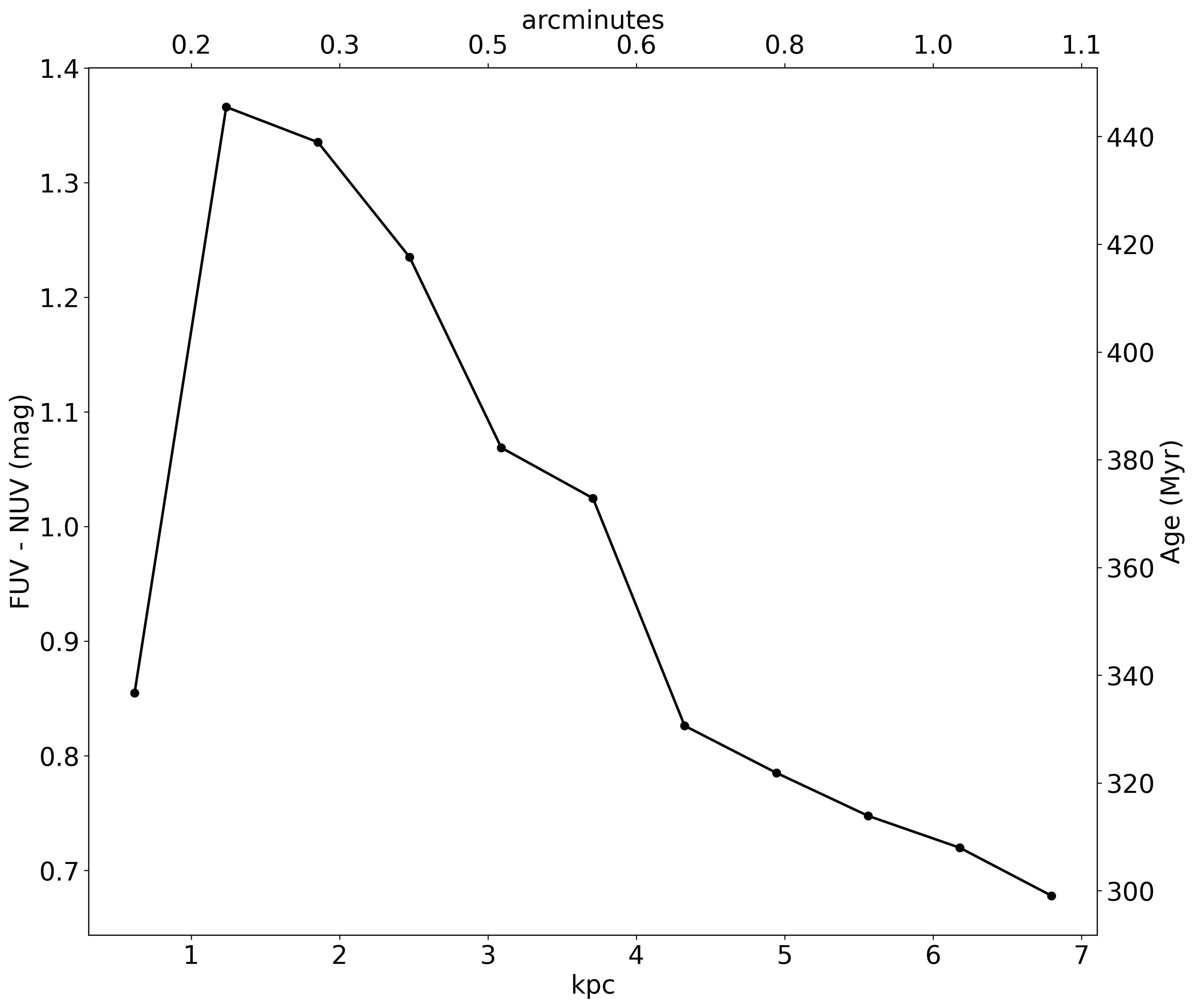}&
     \includegraphics[width=5.7cm,height=5.7cm,keepaspectratio]{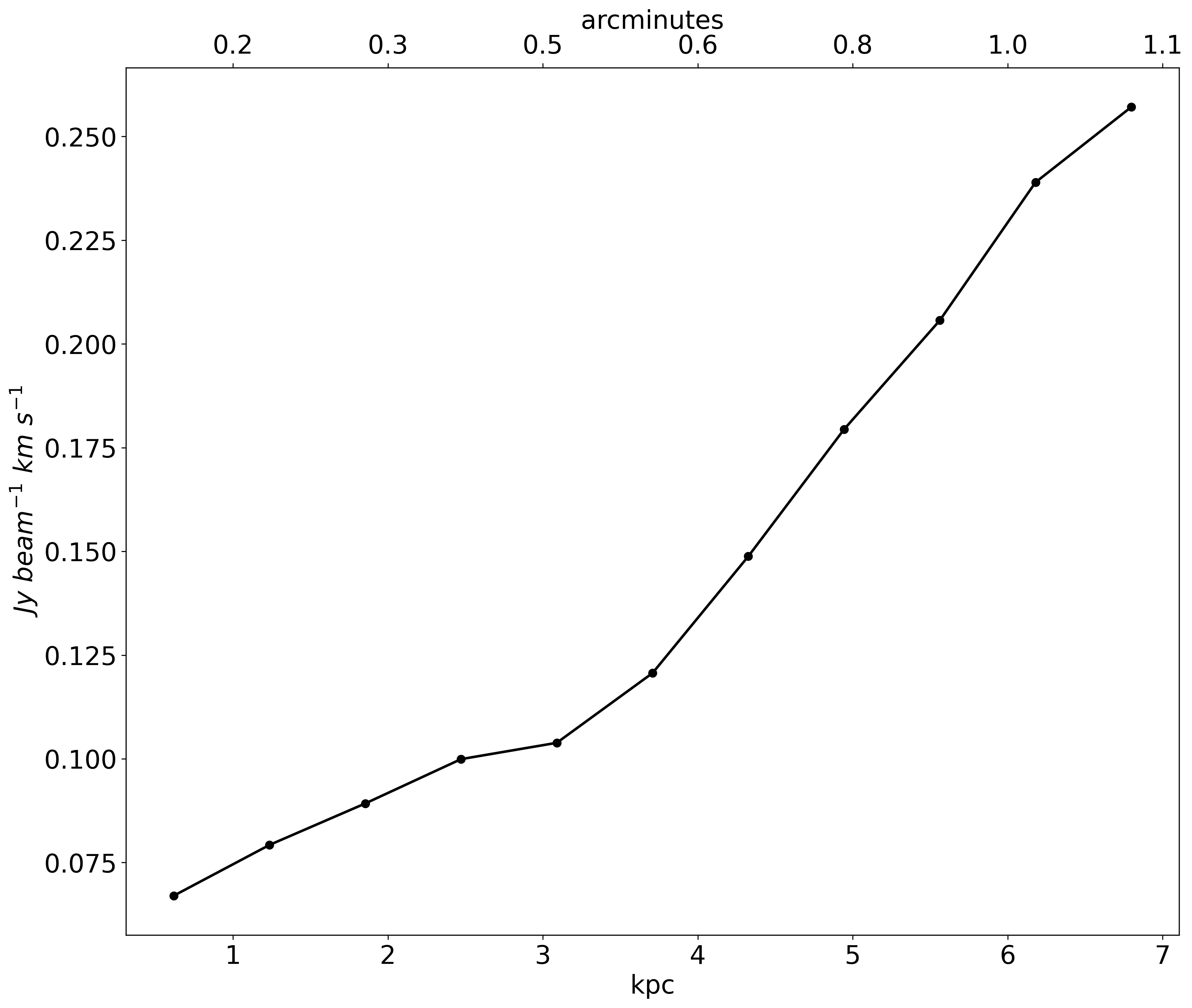}&
      \includegraphics[width=5.7cm,height=5.7cm,keepaspectratio]{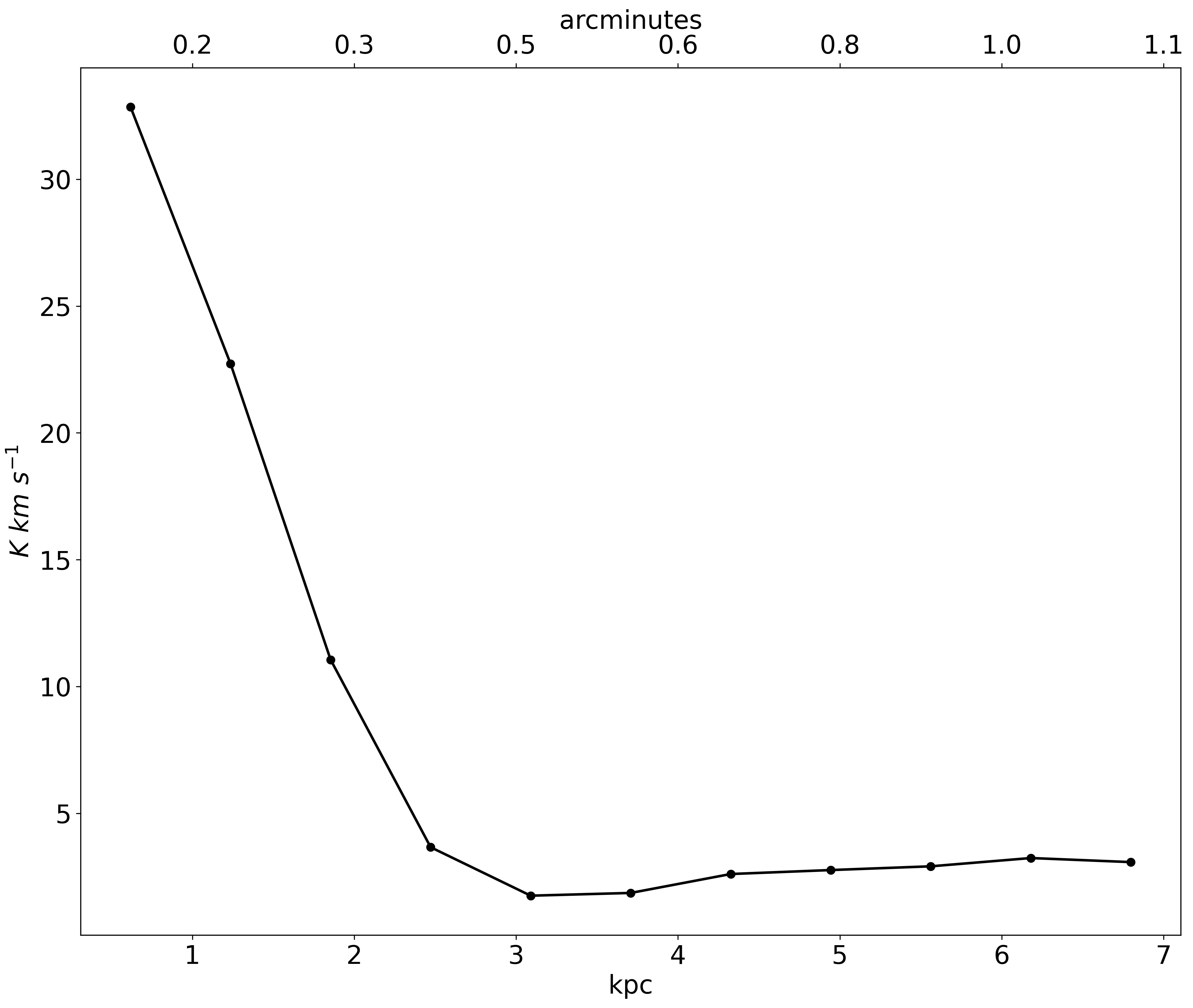}
\end{tabular}
\caption{NGC 4579. First row: 1) $FUV-NUV$ colour map. The black contour shows the stellar bar detected in the Spitzer IRAC 3.6 $\mu$ galaxy image. The pixels are colour-coded in units of $FUV-NUV$ colour. The corresponding SSP equivalent ages are also noted in the colour bar. 2) The HI and CO galaxy maps are shown in the next two columns; the stellar bar is plotted in white. The corresponding azimuthally averaged profile of the galaxy is shown in the bottom row. The main bodies of the galaxies were averaged in elliptical annuli of width 0.1 arcmin (0.62 kpc) taking the ellipticity and position angle of the galaxy into account.}\label{figure:fig4}
\end{figure*}

Figure~\ref{figure:fig4} shows the maps and profiles we created to understand the FUV$-$NUV colour and HI and  H$_{2}$ distribution of NGC 4579. This is very similar to the observations made from NGC 3351: the region along the major axis of the bar hosts stellar populations of age $\geq$ 350 Myr and the nuclear or central sub-kpc region shows an embedded bluer and younger clump of star formation ($\sim$ 150-250 Myr). When we compare the FUV$-$NUV colour map/profile with HI and H$_2$ maps/profiles, we see that the  circular region without the central nuclear region lacks molecular/neutral hydrogen and star formation. The central sub-kpc nuclear region of the galaxy hosts a significant molecular gas content and star formation, but lacks HI. The length of the region that is covered by the stellar bar coincides with the region that is devoid of recent star formation and HI. There is evidence for an AGN in the central region, which can contribute to the UV flux at the centre. The age estimated from the FUV$-$NUV colour in the very central regions can therefor be affected by the AGN.

\subsection{NGC 4725}

NGC 4725 ($\alpha$(J2000) = 12:50:26.6 and $\delta$(J2000) = $+$25:30:03) 
is part of the interacting pair of galaxies NGC 4725 and NGC 4747, which belongs to the spiral-rich Coma I group of galaxies \citep{Gregory_1977,Wevers_1984}. However, we note that this galaxy is marked as a field galaxy in the identification of barred galaxies and environment by \citet{vandenberg_2002}. The galaxy is seen nearly face-on, as shown in Figure~\ref{figure:fig1}. The galaxy is marked as having a peculiar morphology in optical-imaging data. This is a double-barred galaxy with a peculiar long bar that twists sharply along the length of the bar \citep{Erwin_2004,Erwin_2005}. An AGN is located in the central region of the galaxy and is classified as Seyfert 2 \citep{Ho_1997b}. The galaxy also hosts a classical bulge \citep{Fisher_2010}.\\

\begin{figure*}[h]
\centering
\begin{tabular}{ccc}
    \includegraphics[width=5.7cm,height=5.7cm,keepaspectratio]{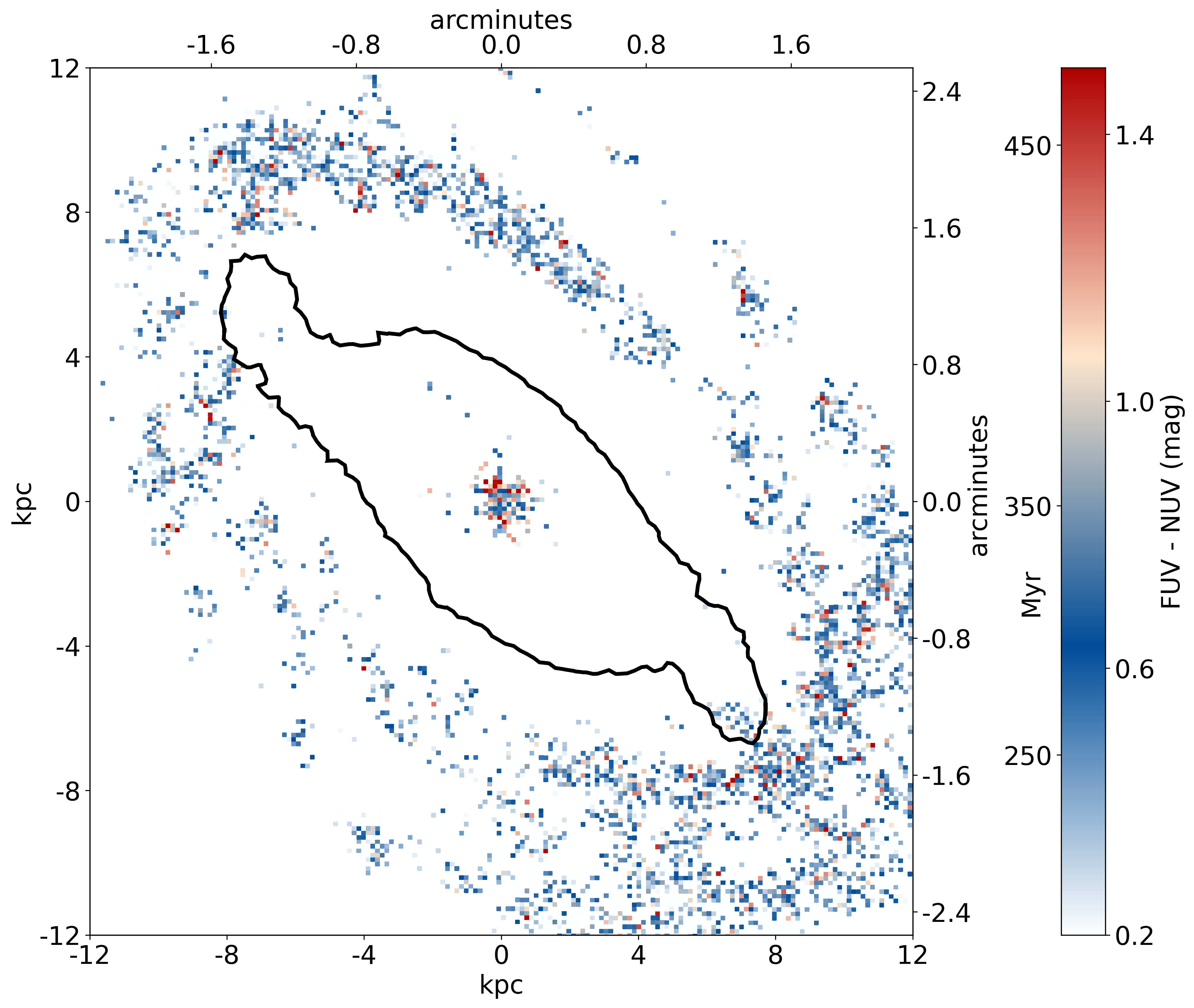}&
    \includegraphics[width=5.7cm,height=5.7cm,keepaspectratio]{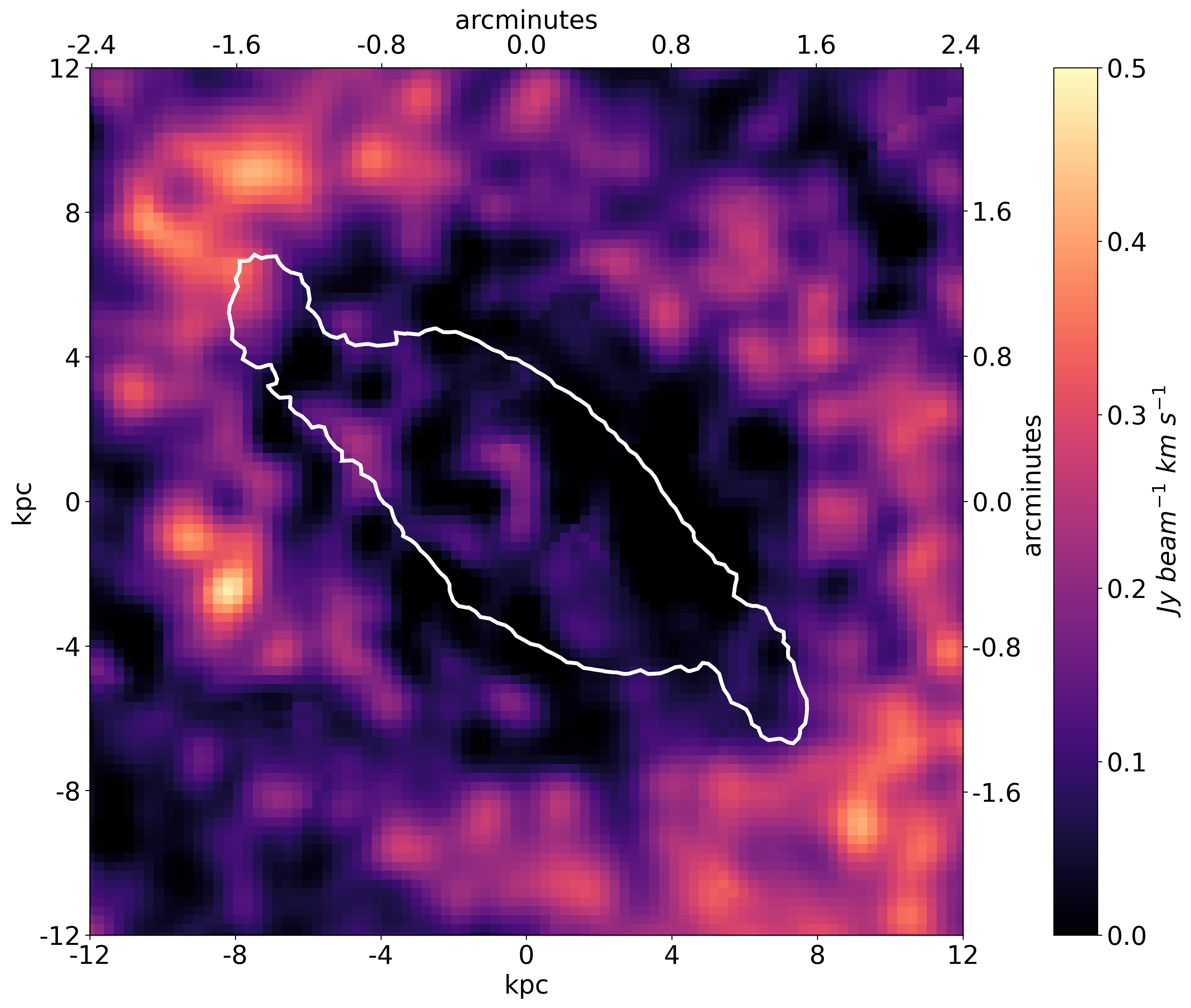}&
    \includegraphics[width=5.7cm,height=5.7cm,keepaspectratio]{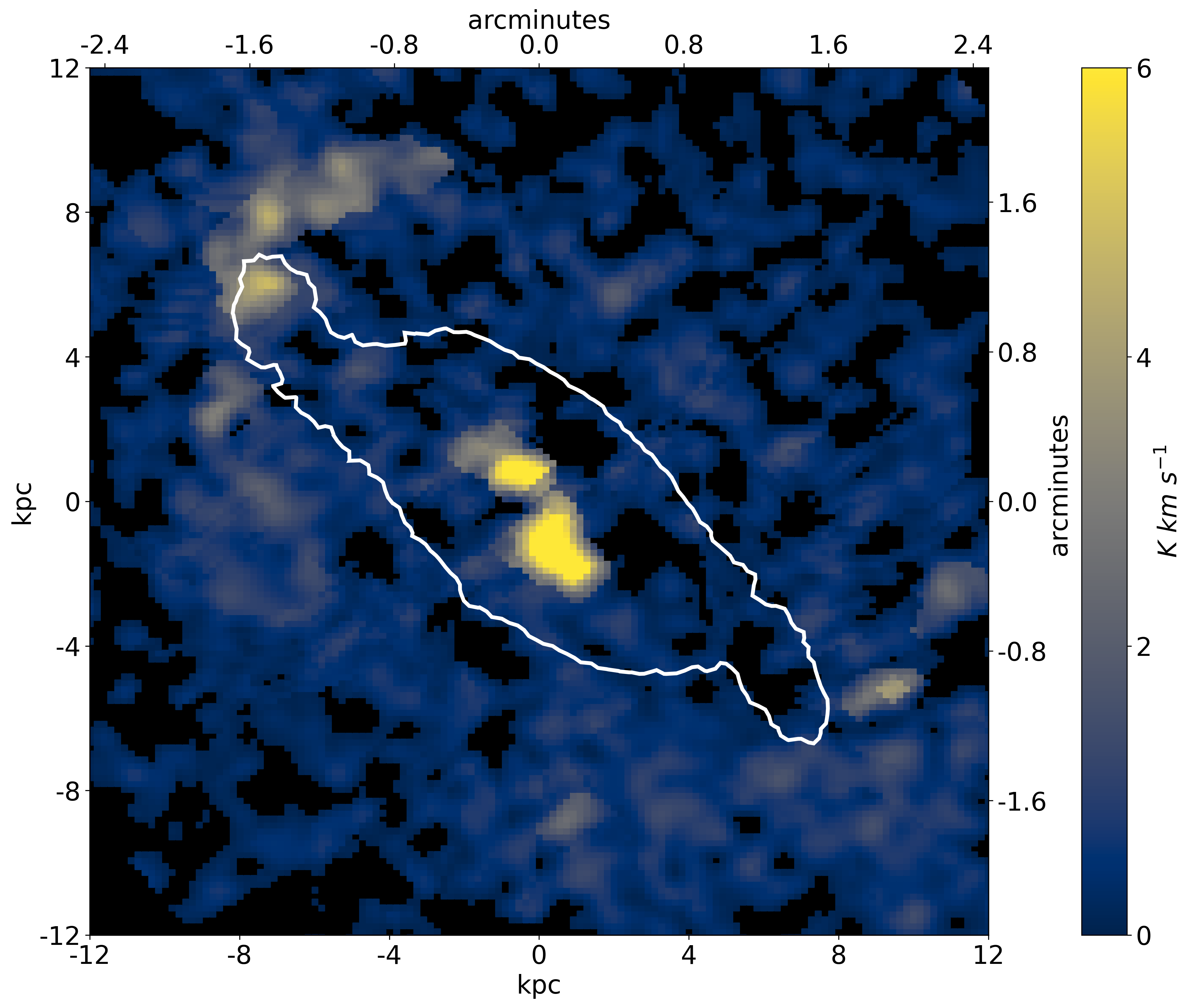}\\[2\tabcolsep]
    \includegraphics[width=5.7cm,height=5.7cm,keepaspectratio]{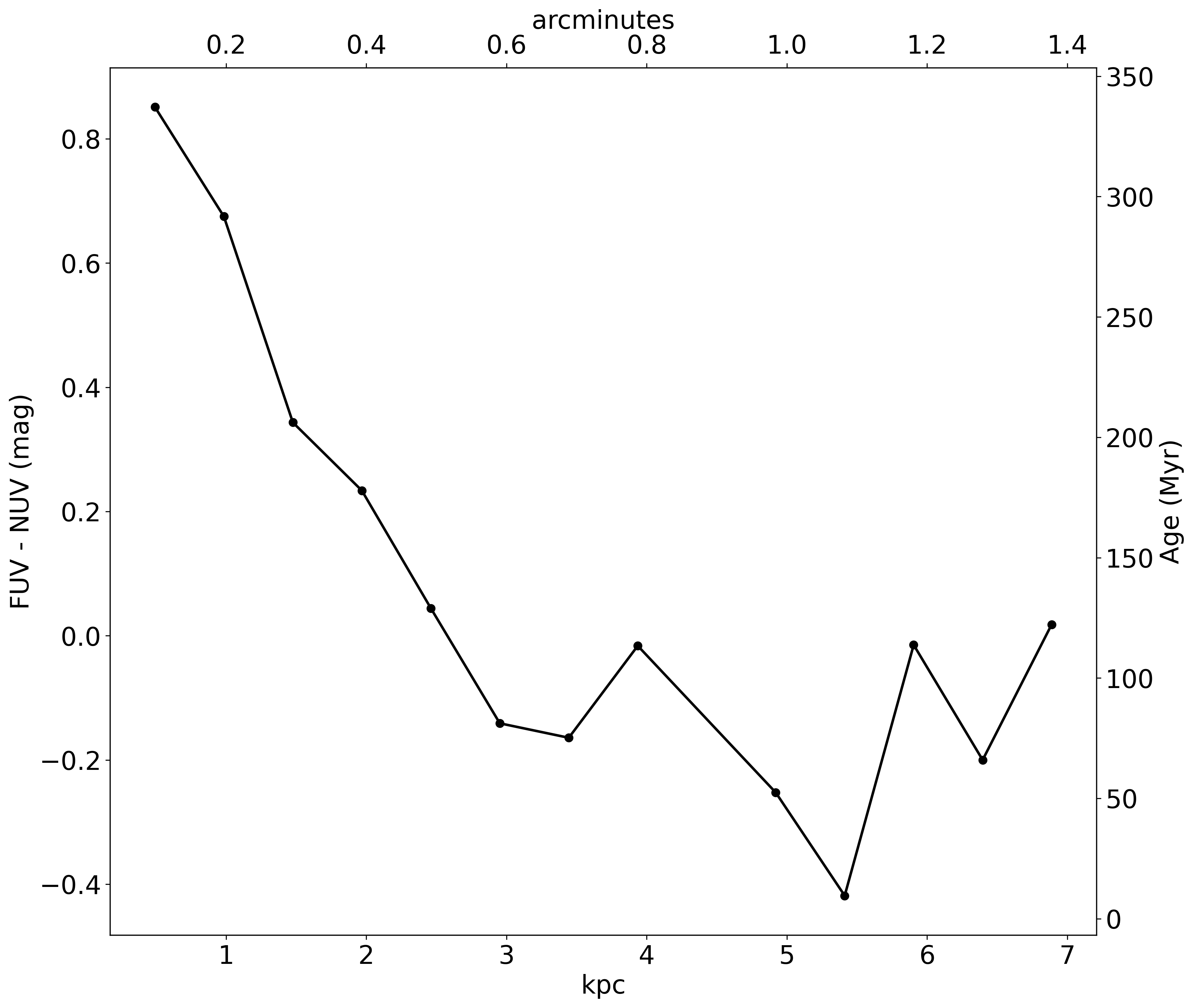}&
     \includegraphics[width=5.7cm,height=5.7cm,keepaspectratio]{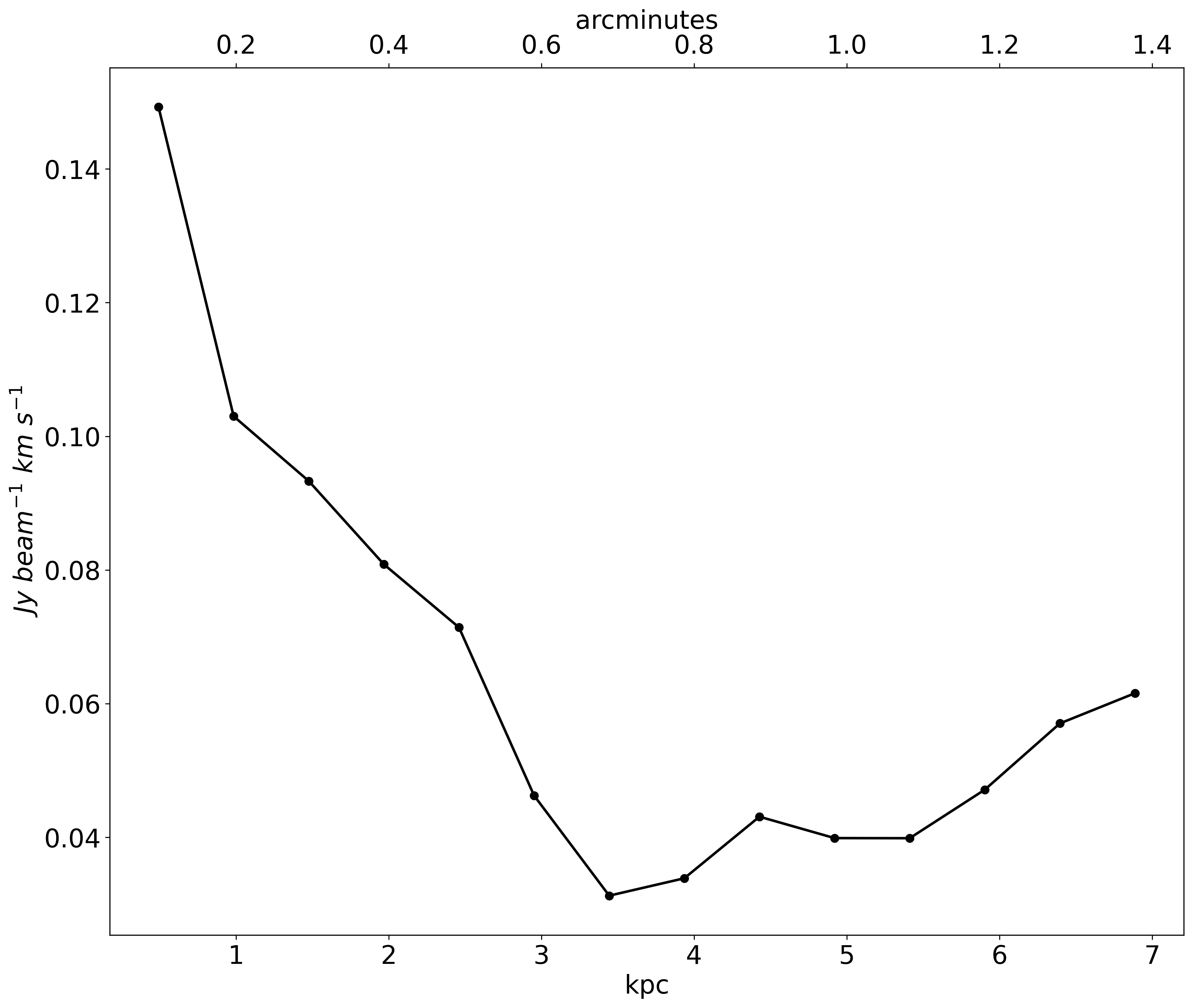}&
      \includegraphics[width=5.7cm,height=5.7cm,keepaspectratio]{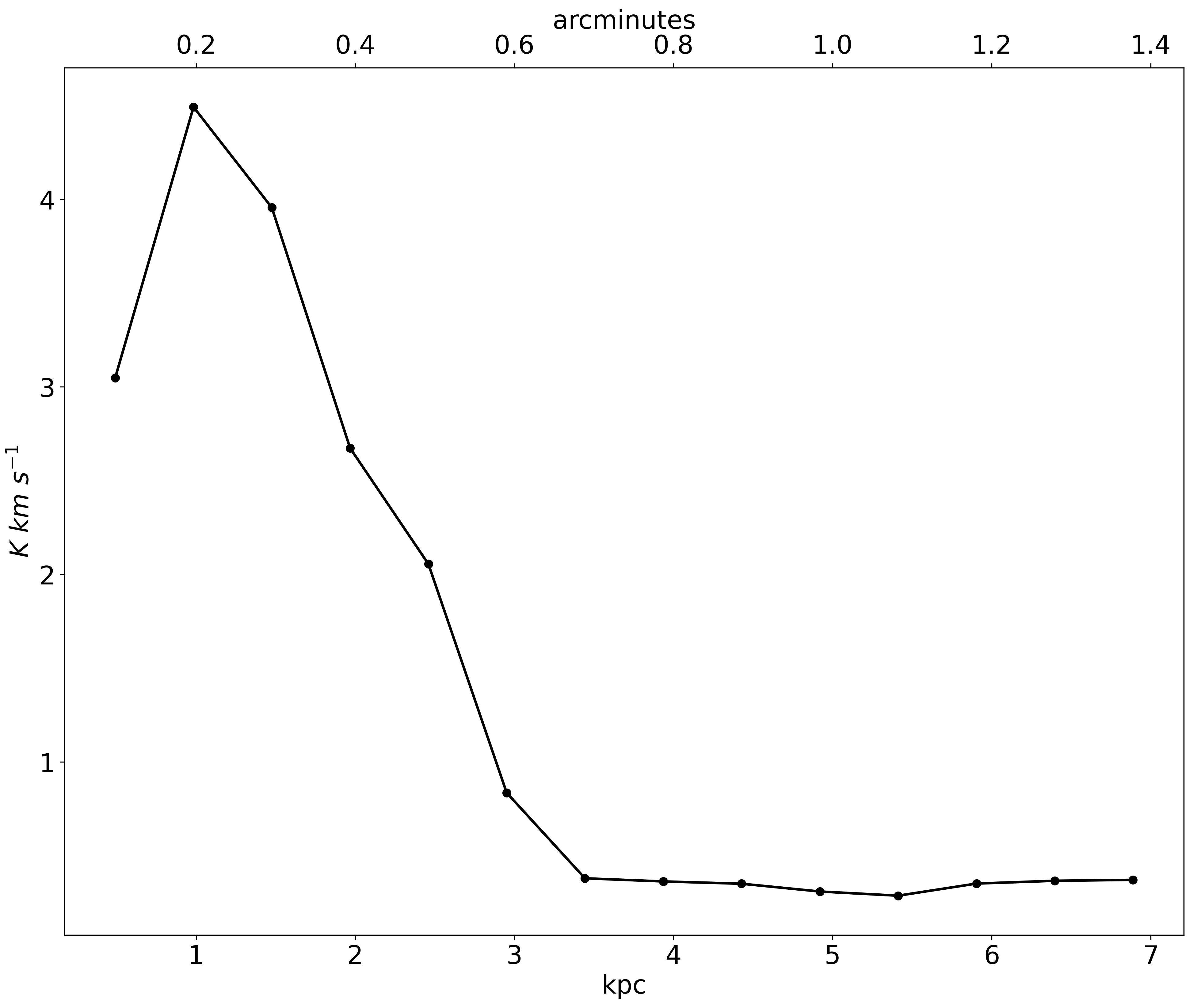}
\end{tabular}
\caption{NGC 4725. First row: 1) $FUV-NUV$ colour map. The black contour shows the stellar bar detected in the Spitzer IRAC 3.6 $\mu$ galaxy image. The pixels are colour-coded in units of $FUV-NUV$ colour. The corresponding SSP equivalent ages are also noted in the colour bar. 2) The HI and CO galaxy maps are shown in the next two columns; the stellar bar is plotted in white. The corresponding azimuthally averaged profile of the galaxy is shown in the bottom row. The main bodies of the galaxies were averaged in elliptical annuli of width 0.1 arcmin (0.5 kpc) taking the ellipticity and position angle of the galaxy into account.}\label{figure:fig5}
\end{figure*}

Figure~\ref{figure:fig5} shows the maps and profiles we created to understand the FUV$-$NUV colour and HI and  H$_{2}$ distribution of NGC 4725. When we compare the FUV$-$NUV colour map and profile with the HI and H$_{2}$ maps and profiles, the circular region without the central nuclear region lacks molecular/neutral hydrogen and star formation. The central sub-kpc nuclear region of the galaxy hosts molecular gas content, little star formation, and HI. The length of the region that is covered by the stellar bar coincides with the region that is devoid of recent star formation and HI. This galaxy also hosts an AGN, which can therefore alter the age estimated from FUV$-$NUV colour in the very central regions.

\section{Discussion}

Stellar bars play a major role in the slow (secular) evolution of disc galaxies as opposed to the much faster effects expected from the large-scale environments within which galaxies reside (see \citealt{Combes_1981,Combes_1990,Debattista_2004,Kormendy_2004}). The enhanced nuclear or central star formation observed in barred spiral galaxies is understood to be due to the result of bar-driven processes that funnel the gas inwards,  which aids in intense nuclear star formation (\citealt{Athanassoula_1992,Ho_1997a,Sheth_2005,Coelho_2011,Ellison_2011,Oh_2012, Wang_2012, Lin_2020, Wang_2020}) and can be responsible for the formation of a pseudo-bulge in disc galaxies \citep{Sanders_1976,Roberts_1979,Athanassoula_1992,Ho_1997a,Kormendy_2004, Jogee_2005, Lin_2017, Spinoso_2017}. \\

However, in the absence of an external gas supply, to counter the accumulation of gas (both neutral and  H$_{2}$) in the central few kpc, regions in the galaxy disc might be devoid of gas. Such regions are indeed observed in the form of cavities (or holes) in HI within the the co-rotation radius of a few barred disc galaxies \citep{Consolandi_2017,George_2019a,Newnham_2020}. The region outside the co-rotation radius can have significant HI, so that even when the central regions are devoid of star formation, quenched barred galaxies can still be gas rich \citep{George_2019b}. This observed scenario is consistent with recent simulations in which the gas surface density is expected to be reduced along the bar region near the periphery of the co-rotation radius. The snapshots of regions around the bar in gas and stellar surface density maps obtained from simulations are worth exploring. We refer in particular to the results from simulations of \citet{Gavazzi_2015}, \citet{Spinoso_2017}, and \citet{Donohoe-Keyes_2019}, in which a region within the bar co-rotation radius has a reduced gas surface density and is observed to have no recent star formation but is composed of an old stellar population. This region is very similar to the star formation desert reported by \citet{James_2009},\citet{James_2016} and \citet{James_2018}, kpc-scale region dominated by line emission not linked to AGN or star formation, but instead attributed to post-asymptotic giant branch stars \citep{James_2015,Percival_2020}.\\

The multi-wavelength study of four barred galaxies using the archival data ranging from the UV to the optical, infrared, HI, and  H$_{2}$ (as traced by CO) showed that star formation quenching occurs in the bar region for three galaxies. There is no star formation in the last 100-200 Myr within the bar co-rotation radius (except at the central sub-kpc regions), as is evident from the FUV$-$NUV colour maps of NGC 3351, NGC 4579, and NGC 4725. For these galaxies, the  region hosts stellar populations older than $\sim$ 400 Myr, and the corresponding error in age is mostly smaller than $\sim$ 150 Myr (ranges between $\sim$ 1 - 150 Myr). Therefore, considering these errors will still signify no star formation in the last $\sim$ 250 Myr in the bar co-rotation radius.

This region also is devoid of HI and H$_{2}$ at the detection limit of observations based on VLA and IRAM. The lack of H$_{2}$ and HI in this region implies that the stellar bar might have redistributed the gas. The stellar bar can funnel the gas to the  centre and can be the reason for the significant molecular gas content and recent star formation observed in the central sub-kpc nuclear or central region. This funneling of gas to the central sub-kpc region would have depleted the gas in the bar-region and thereby suppressed star formation through lack of fuel. On the other hand, significant amount of HI is present outside the length of bar along with a young stellar population. The absence of H$_2$ and HI in the bar region of three galaxies can be considered as support of the scenario of gas redistribution.\\

\subsection{Special case of NGC 2903}

The star formation progression and HI/H$_{2}$ distribution of NGC 2903 differs from that of the other three galaxies studied here. Although a prominent stellar bar is detected in Spitzer 3.6 $\mu$ imaging data, NGC 2903 shows recent star formation (100-300 Myr) and significant HI and H$_2$ along the bar co-rotation radius. This galaxy is located in an isolated environment and is an ideal case to examine the secular evolution due to bar-induced processes. Based on the distribution of the HCN(1-0) emission along the stellar bar of NGC 2903 and on a comparison with results from numerical simulations, it was shown that the bar can be relatively young, with an age between 200-600 Myr \citep{Martin_1997,Leon_2008}. We speculate about a scenario in which the stellar bar is yet to completely redistribute the gas from the co-rotation radius to the central few kpc region of the galaxy. This is supported by the presence of H$_{2}$ along the length of the bar, and the region around the bar has significant star formation. We furthermore searched for signatures of such an observed feature in simulations and found that it is indeed the case, where a recently formed bar takes 1-2 Gyr to evacuate the gas from the region around the bar \citep{Donohoe-Keyes_2019}. We note that NGC 2903 belongs to the field environment and all other galaxies are part of a loose group or cluster of galaxies. The role of environment in building the bars in disc galaxies is still widely debated \citep{vandenberg_2002,Lin_2014}. Other mechanisms might quench star formation in dense environments.\\

\subsection{Possible caveats}

The extinction due to dust internal to the galaxies studied here can absorb the UV photons from the regions covered by the stellar bar, which can then be re-emitted in the infrared \citep{Silva_1998}. We do not have an independent extinction measurement for each galaxy and therefore assumed an arbitrary extinction value $A_{V}$=0.2 for the galaxies and adopted the  \citet{cardelli_1989} relation, scaling the $A_{V}$ value to $A_{FUV}$ and $A_{NUV}$ for two different $R_{V}$ values (MilkyWay: 3.1 and LMC: 3.3). In both cases we obtained a difference in extinction between FUV and NUV ($A_{FUV}$ - $A_{NUV}$) = -0.07. This implies that if we were to adopt an arbitrary extinction value, the $FUV-NUV$ colour would become redder by 0.07 mag. The ages would accordingly become older if we were to consider additional extinction internal to the galaxy. We also note that the corresponding change in age for 0.07 mag change in colour is negligible and the nature of the colour profiles that we  showed for each galaxy will remain the same. The detailed radiative transfer modelling of NGC 3351 by \citet{Nersesian_2020} showed that the dust content and hence the attenuation along the bar region is negligible. The Spitzer 3.6 ${\mu}$ infrared imaging data of our galaxies show a similar pattern of star formation progression in the region that is covered by the stellar bar.  The 3.6 ${\mu}$ emission is an extinction-free tracer for the cool stars that dominate the underlying stellar mass \citep{Meidt_2014}.

Other mechanisms, such as AGN feedback and morphological quenching aided by bulges, might be responsible for the quenching of star formation in the inner or central regions of our sample galaxies. All of the four barred galaxies presented here host bulges: NGC 2903, NGC 335, and NGC 4579 have a pseudo-bulge, and NGC 4725 has a classical bulge \citep{Fisher_2008}. The enhancement of star formation in the central region of barred galaxies (prior to  bar quenching) can build-up pseudo-bulges \citep{Kormendy_2004}, and this may be the case that we see in  NGC 2903, NGC 3351, and NGC 4579. Morphological quenching aided by bulges is generally driven by classical bulges, which  are formed during major  merger events \citep{Martig_2009}. There can be a combined effect of morphological quenching and bar quenching in NGC 4725. Moreover, NGC 4725 and NGC 4579 host AGN in their centres \citep{Ho_1997a,Ho_1997b}. The quenching due to mechanisms other than bar quenching can also operate in these galaxies and cannot be completely ruled out.

Dynamical effects due to the action of the stellar bar can affect the observations presented here through the redistribution of stars from the very location at which they are born \citep{Renaud_2015}. The very young stellar population found in the central regions of two galaxies (NGC 4579 and NGC 3351) can then be formed elsewhere, but now decoupled and moved to the very central region by the action of stellar bar.\\

We presented evidence for gas redistribution due to the stellar bar and subsequent star formation quenching within the bar co-rotation radius in three barred galaxies. The main result of our analysis is a region between the nuclear region and the end of the bar that is devoid of gas and star formation in the past a few 100 Myr. Star formation is quenched in this region, and the absence of  molecular/neutral hydrogen gas implies that no further star formation is possible, or in other words, bar quenching can be a dominant star formation suppression mechanism in the three galaxies. In the absence of an external supply of gas, the star formation in the centre will deplete the  H$_{2}$ completely and the galaxy will eventually be devoid of star formation in the bar and the central nuclear region. Bar quenching in galaxies belonging to the dense environments of groups/clusters can quench the galaxies faster with no further supply of gas. Future high spatial resolution multi-wavelength observations of a large number of barred galaxies with facilities such as Atacama Large Millimeter/submillimeter Array (ALMA), the VLA, and the Very Large Telescope (VLT) along with UV observations with the Hubble Space Telescope (HST) can give further insights into the gas kinematics along the co-rotation radius.\\

Bar quenching can be a dominant mechanism in the quenching of disc galaxies at high redshifts. Barred disc galaxies are known to exists from redshift $\sim$ 1 \citep{Sheth_2008}, and there is evidence for even redshift $\sim$ 3 observations of bars in galaxies \citep{Hodge_2019}. In a scenario in which bar quenching  occurs in these high-redshift galaxies (after discs formation is stabilised and there is time to form bars), a mechanism exists to convert star-forming galaxies into passive galaxies without invoking major morphology changes. This also implies that no other quenching mechanisms such as AGN feedback are required to quench star formation in isolated disc galaxies in which the environment plays no role. Future high spatial resolution imaging observations with the JWST and the ELT (TMT, GMT, and the ELT) might detect such galaxies that undergo bar quenching at high redshifts.

\section{Summary}

The possibility that stellar bars might quench star formation within the co-rotation radius of barred galaxies was studied here using four barred galaxies drawn from HERACLES survey. Based on multi-wavelength data, we demonstrated that the regions near to the bar for three galaxies (NGC 3351, NGC 4579, NGC 4725) are devoid of any ongoing/recent star formation and that this same region is devoid of molecular and neutral hydrogen. The central sub-kpc region of all galaxies hosts an abundant supply of  H$_{2}$ , and the region along the bar is devoid of neutral and  H$_{2}$. However, a significant amount of neutral hydrogen is present outside the stellar bar region. This observation can be considered as direct evidence  that the stellar bar in galaxies dynamically redistributes the gas and depletes the region close to the bar of fuel for star formation. For one galaxy (NGC 2903) the co-rotation radius of the bar is not devoid of star formation and there is an abundant supply of  H$_{2}$ throughout the bar region. This can be due to the action of bar that is yet to fully redistribute the gas.

\begin{acknowledgements}
We thank the anonymous referee for the comments, which improved the scientific content of the paper. We thank Adam Leroy for providing VLA HI data on NGC 4579 and NGC 4725. SS acknowledges support from the Science and Engineering Research Board, India through a Ramanujan Fellowship. This work made use of THINGS, 'The HI Nearby Galaxy Survey' \citep{Walter_2008}. This work made use of HERACLES, `The HERA CO-Line Extragalactic Survey’ \citep{Leroy_2009}. We acknowledge the usage of the HyperLeda database (http://leda.univ-lyon1.fr). This research made use of Astropy, a community-developed core Python package for Astronomy \citep{Astropy_Collaboration_2018}. This research made use of Montage. It is funded by the National Science Foundation under Grant Number ACI-1440620, and was previously funded by the National Aeronautics and Space Administration's Earth Science Technology Office, Computation Technologies Project, under Cooperative Agreement Number NCC5-626 between NASA and the California Institute of Technology. Funding for SDSS-III has been provided by the Alfred P. Sloan Foundation, the Participating Institutions, the National Science Foundation, and the U.S. Department of Energy Office of Science. The SDSS-III web site is http://www.sdss3.org/. SDSS-III is managed by the Astrophysical Research Consortium for the Participating Institutions of the SDSS-III Collaboration including the University of Arizona, the Brazilian Participation Group, Brookhaven National Laboratory, Carnegie Mellon University, University of Florida, the French Participation Group, the German Participation Group, Harvard University, the Instituto de Astrofisica de Canarias, the Michigan State/Notre Dame/JINA Participation Group, Johns Hopkins University, Lawrence Berkeley National Laboratory, Max Planck Institute for Astrophysics, Max Planck Institute for Extraterrestrial Physics, New Mexico State University, New York University, Ohio State University, Pennsylvania State University, University of Portsmouth, Princeton University, the Spanish Participation Group, University of Tokyo, University of Utah, Vanderbilt University, University of Virginia, University of Washington, and Yale University.
\end{acknowledgements}


\end{document}